\documentclass[]{aa}

\usepackage{graphicx}
\usepackage{txfonts}
\usepackage{natbib,twoopt}
\usepackage[breaklinks=true]{hyperref} 
\usepackage{caption} 
\bibpunct{(}{)}{;}{a}{}{,}             
\makeatletter
  \newcommandtwoopt{\citeads}[3][][]{\href{http://adsabs.harvard.edu/abs/#3}%
    {\def\hyper@linkstart##1##2{}%
     \let\hyper@linkend\@empty\citealp[#1][#2]{#3}}}
  \newcommandtwoopt{\citepads}[3][][]{\href{http://adsabs.harvard.edu/abs/#3}%
    {\def\hyper@linkstart##1##2{}%
     \let\hyper@linkend\@empty\citep[#1][#2]{#3}}}
  \newcommandtwoopt{\citetads}[3][][]{\href{http://adsabs.harvard.edu/abs/#3}%
    {\def\hyper@linkstart##1##2{}%
     \let\hyper@linkend\@empty\citet[#1][#2]{#3}}}
  \newcommandtwoopt{\citeyearads}[3][][]%
    {\href{http://adsabs.harvard.edu/abs/#3}
    {\def\hyper@linkstart##1##2{}%
     \let\hyper@linkend\@empty\citeyear[#1][#2]{#3}}}
\makeatother

\begin{document}

\title{New quasars behind the Magellanic Clouds. II. Spectroscopic confirmation
of 136 near-infrared selected candidates\footnote{Tables 1, 2, 4 and B.1 are
available in electronic form at the CDS via anonymous ftp to cdsarc.u-strasbg.fr
(130.79.128.5) or via http://cdsweb.u-strasbg.fr/cgi-bin/qcat?J/A+A/.}}

\author{
Valentin D. Ivanov\inst{1}
\and
Maria-Rosa L. Cioni\inst{2}
\and
Michel Dennefeld\inst{3}
\and
Richard de Grijs\inst{4,5,6}
\and
Jessica E. M. Craig\inst{7}
\and
Jacco Th. van Loon\inst{7}
\and
Clara Pennock\inst{7,8}
\and
Chandreyee Maitra\inst{9}
\and
Frank Haberl\inst{9}
}

\offprints{V. Ivanov, \email{vivanov@eso.org}}

\institute{
European Southern Observatory, Karl-Schwarzschild-Str. 2, 
D-85748 Garching bei M\"unchen, Germany
\and 
Leibniz-Institut f\"ur Astrophysik Potsdam, An der Sternwarte 16, 
D-14482 Potsdam, Germany
\and
Institut d'Astrophysique de Paris (IAP), 98bis Bd Arago, 75014, 
Paris, France
\and
School of Mathematical and Physical Sciences, Macquarie University,
Balaclava Road, Sydney, NSW 2109, Australia
\and
Astrophysics and Space Technologies Research Centre, Macquarie
University, Balaclava Road, Sydney, NSW 2109, Australia
\and
International Space Science Institute–-Beijing, 1 Nanertiao,
Zhongguancun, Hai Dian District, Beijing 100190, China
\and
Lennard-Jones Laboratories, Keele University, ST5 5BG, UK
\and
Institute for Astronomy, University of Edinburgh, Royal
Observatory, Edinburgh EH9 3HJ, UK
\and
Max-Planck-Institut f\"ur extraterrestrische Physik,
Giessenbachstrasse 1, 85748 Garching, Germany
}

\date{Received 2 November 1002 / Accepted 7 January 3003}

\abstract 
{Quasi--stellar objects (QSOs) are a basis for an absolute 
reference system for astrometric studies. There is a need for 
creating such system behind nearby galaxies, to facilitate the 
measuring of the proper motions of these galaxies. However, the 
foreground contamination from the galaxies themselves is a 
problem for the QSO identification.}
{We search for new QSOs behind both Magellanic Clouds, the 
Magellanic Bridge, and the Magellanic Stream.}
{We identify QSO candidates with a combination of near--infrared 
colors and variability criteria from the public ESO Visual and 
Infrared Survey Telescope for Astronomy (VISTA) Magellanic 
Clouds (VMC) survey. We confirm their nature from broad emission 
lines with low-resolution optical spectroscopy.}
{We confirmed the QSO nature of 136 objects. They are
distributed as follows: 12 behind the LMC, 37 behind the SMC, 63
behind the Bridge, and 24 behind the Stream. The QSOs span a
redshift range from $z$$\sim$0.1 to $z$$\sim$2.9. A comparison
of our quasar selection with the Quaia quasar catalog, based on
{\it Gaia} low-resolution spectra, yields a selection and
confirmation success rate of 6-19\,\%, depending on whether the
quality of the photometry, the magnitude ranges and the colors
are considered. Our candidate list is rather incomplete, but the
objects in it are likely to be confirmed as quasars with
$\sim$90\,\% probability. Finally, we report a list of 3609
objects across the entire VMC survey that match our color and
variability selection criteria; only 1249 of them have {\it Gaia}
counterparts.}
{Our combined infrared color and variability criteria for QSO 
selection prove to be efficient -- $\sim$90\,\% of the observed
candidates are bona fide QSOs and allow to generate a list
of new high-probability quasar candidates.
}
\authorrunning{V. Ivanov et al.}
\titlerunning{New QSOs behind the LMC and SMC. II.}

\maketitle

\section{Introduction}\label{sec:intro}

Quasi--stellar objects (QSOs) are distant galaxies with central 
supermassive black holes accreting from their surroundings. They
are only luminous on a scale of a few hundred parsecs, therefore
they appear as point-like objects.

Quasar candidates are often selected based on optical or infrared 
(IR) variability 
\citep{1994MNRAS.268..305H,2023A&A...674A..24C,2023MNRAS.525.5795T} 
radio emission
\citep{1996AJ....112..407G,2015MNRAS.449.2818T} 
X-ray 
\citep{1991Natur.353..315S,2023arXiv230600961W} 
or mid--IR colors 
\citep{2004ApJS..154..166L,2022Univ....8..356S}. 
Multi-wavelength selections are also used 
\citep{2015MNRAS.452.3124D,2019A&A...630A.146S}.
Bona-fide QSOs are confirmed by their broad emission lines (although
their absence does not imply the object is not a QSO),
that also help to derive their redshifts 
\citep{2001AJ....122..549V}. Millions of such objects are known today
\citep{2023arXiv230801505F,2023arXiv230801507F,2023arXiv230617749S,2023PASA...40...10O}
and the lists continue to be updated with the latest surveys.

Quasars provide insights into the structure and evolution of galaxies and
are cosmological probes of the intervening interstellar medium. They 
can also serve as an absolute astrometric reference frame as distant 
unmoving point-source
objects -- a necessity for measuring the small proper motions (PMs) 
within nearby galaxies. However, it is challenging to identify QSOs 
behind these galaxies because of the rich and diverse stellar content 
of the galaxies themselves that blur the boundaries of the stellar 
locus. Furthermore, the internal reddening inside the foreground 
galaxies modifies the QSOs' colors, especially the optical ones. 

Finding QSOs is difficult behind the nearby Magellanic Clouds system
because of the foreground contamination and in particular behind the
Small Magellanic Cloud (SMC), because it exhibits a notable depth and
associated with it stronger source confusion
\citep{2015AJ....149..179D}. Nevertheless, a significant number of 
QSOs have been identified there by studies that have specifically
targeted this region
\citep{1986PASP...98..635B,
2002ApJ...569L..15D,2003AJ....125.1330D,2003AJ....126..734D,2005A&A...442..495D,
2003AJ....125....1G,
2009ApJ...701..508K,
2012ApJ...746...27K,2011ApJS..194...22K,
2010A&A...518A..10V,
2013ApJ...775...92K}. 

Optical searches can confuse QSOs with stars
\citep{2015MNRAS.453.2341V,2022MNRAS.515.6046P} or miss more highly
obscured QSOs. Moving to radio \citep{2021MNRAS.506.3540P} or near--
and mid--IR wavebands can help to alleviate the latter problem.
This prompted us to search for QSOs in the VISTA \citep[Visual and 
Infrared Survey Telescope for Astronomy;][]{2006Msngr.126...41E} 
Survey of the Magellanic Clouds system \citep[VMC;][]{2011A&A...527A.116C}. 
Until now the largest single spectroscopically confirmed sample of 
quasars -- 758 QSOs -- was an OGLE \citep[Optical Gravitational 
Lensing Experiment][]{1992AcA....42..253U} follow up by 
\citet{2013ApJ...775...92K} and more recently -- by the Quaia
survey based on the low-resolution (R$\sim$30) {\it Gaia}
spectra \citep{2023arXiv230617749S}. Our main goal is to increase
the number of known quasars behind the Large and the Small
Magellanic Clouds, aiming to improve the proper motion reference
system. We reported our first results -- 37 (34 new) QSOs, also
spectroscopically confirmed -- in \citet{2016A&A...588A..93I}.
This is the second paper in our series, with further spectroscopy
from VLT (Very Large Telescope) and SAAO (South African Astronomical 
Observatory).

\section{Sample selection}\label{sec:sample}

\subsection{VMC}

The VMC is an ESO public survey with a footprint of 184\degr$^2$ on
the sky around the Large Magellanic Cloud (LMC), SMC, the Magellanic 
Bridge and Stream. The photometry reaches signal--to--noise 
S/N$\sim$10 at 
$K_\mathrm{s}$=20.3\,mag (Vega system) in three epochs in the $Y$ 
and $J$ bands, and in 12 epochs in the $K_\mathrm{s}$ band. More 
epochs are available for some tiles\footnote{Tiles are contiguous 
images that combine six pawprints taken in an offset pattern; 
pawprint is an image obtained on an individual VIRCAM pointing that 
generates an image with gaps between the detectors. See 
\citet{2011A&A...527A.116C} for a description of the VMC observing 
strategy.} and in the tile overlapping areas. The time series 
span nearly 8 years. The VMC science is diverse and includes: star
formation and star formation history, galaxy structure, star
clusters, various types of stars (e.g., RR Lyrae and Cepheids),
proper motions, background galaxies and even distant quasars.

The VMC is carried out at VISTA -- a 4.1m telescope located on Cerro 
Paranal, equipped with VIRCAM 
\citep[VISTA InfraRed CAMera;][]{2006SPIE.6269E..0XD} --
a near--IR wide--field $\sim$1$\times$1.5\degr$^2$) camera.
The data are reduced with the VISTA Data Flow System 
\citep[VDFS;][]{2004SPIE.5493..411I,2004SPIE.5493..401E} at the 
Cambridge Astronomical Survey 
Unit\footnote{\url{http://casu.ast.cam.ac.uk/}}. Their 
photometric calibration is described in
\citet{2018MNRAS.474.5459G}.
The data products are available at the ESO Science Archive 
Facility\footnote{\url{http://archive.eso.org/cms.html}} or at the 
VISTA Science Archive 
\citep[VSA;][]{2012A&A...548A.119C}\footnote{\url{http://archive.eso.org/cms.html}}.

\subsection{Quasar candidate selection}

\begin{figure} 
\centering
\includegraphics[width=8.8cm]{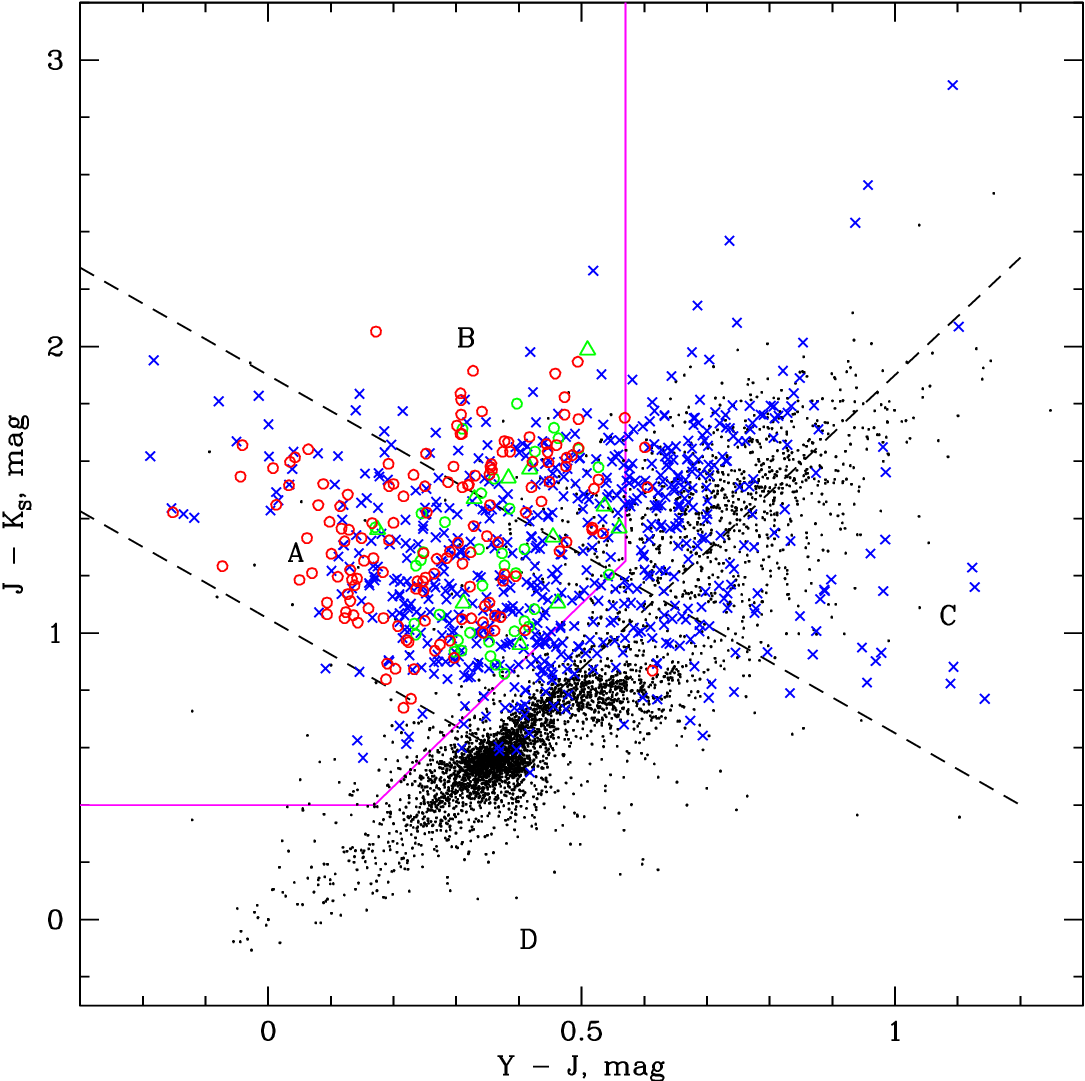} \\
\caption{Color--color diagram showing the color selection of 
candidates. The dashed lines identify regions (labeled with letters) 
with known QSOs and the solid magenta line marks the blue border
of the planetary nebulae locus \citep{2013A&A...549A..29C}. The 
confirmed QSOs from this work are red open circles. The open green 
circles and triangles mark QSOs and non-QSOs from 
\citet{2016A&A...588A..93I}, respectively. Blue $\times$'s indicate 
VMC counterparts to the spectroscopically confirmed QSOs from 
\citet[][VMC photometry selected within a matching radius of 
1\arcsec]{2013ApJ...775...92K}. Black dots are randomly drawn LMC
objects (with errors in all three bands $<$$0.1$\,mag) to mark the 
main stellar locus; some among these in regions B and C are 
contaminating background galaxies.}\label{fig:CCD}
\end{figure}

\begin{figure}
\centering
\includegraphics[width=9.0cm]{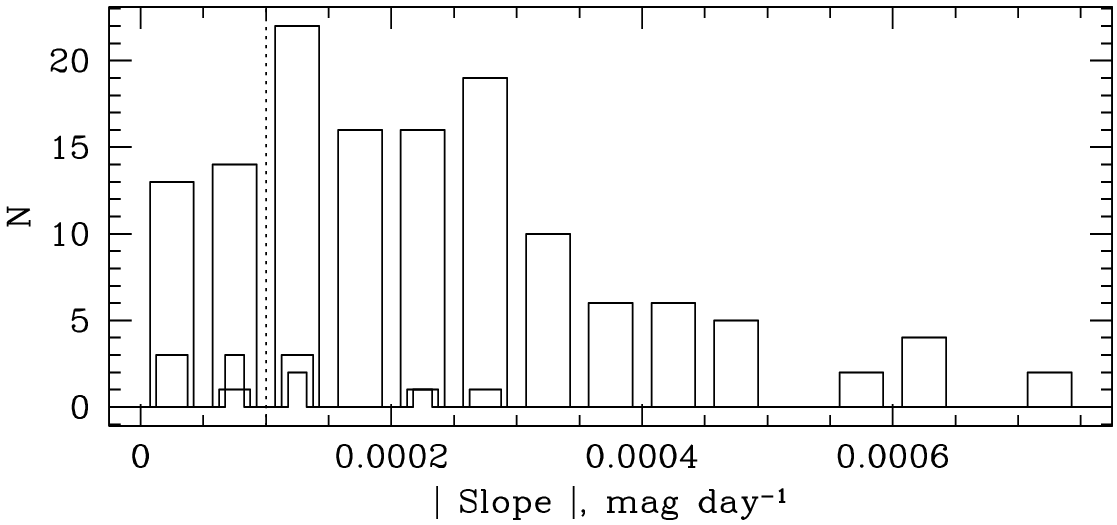}
\caption{Histogram of the absolute slopes from linear fit to the light
curves of the objects in this paper: classified as QSOs (wide bins),
remaining unknown because of poor quality spectra or lack of prominent
emission lines (mid-width bins) and confirmed stars (narrow bins). The
vertical dotted line is the QSO selection limit, defined by
\citet{2013A&A...549A..29C}.}\label{fig:slopes}
\end{figure}

Criteria to identify QSO candidates from VMC were defined by 
\citet{2013A&A...549A..29C} from $117$ known QSOs in a ($Y$$-$$J$)
versus ($J$$-$$K_\mathrm{s}$) color--color diagram (Fig.\ref{fig:CCD})
and from their $K_\mathrm{s}$-band variability, requiring that the
absolute value of the light curve slope exceeds a certain limit:
|Slope|$>$10$^{-4}$\,mag\,day$^{-1}$.
The light curve slope was determined from a simple linear fit of the
$K_\mathrm{s}$ band versus the time of observations in days and the
adopted limit was based on the behavior of the same known quasars.
The quasars were selected with the data available at the time of
Phase 2 submission (DR6 plus some additional images, typically one
per tile). Since then, some data were reprocessed and more
observations became available. An inspection of updated light curves
and slope measurements from them (Fig.\,\ref{fig:slopes}) showed
that a significant number of the spectroscopically confirmed quasars
fall below of the adopted limit which brings up the need for adopting
a more stringently defined criterion in the future, for example based
on damped random walk \citep{2009ApJ...698..895K}.

Here, as in \citet{2016A&A...588A..93I}, we apply the same criteria 
on 18 tiles, mostly peripheral: 3 in LMC, 7 in SMC, 6
in the Bridge and 2 in the Stream. We selected 142 candidates: 15 
in LMC, 40 in SMC, 63 in the Bridge and 24 in the Stream. The Bridge 
and the Stream tiles yield more candidates per tile because of the 
lower foreground contamination from the Magellanic Clouds and the
smaller extinction.
The observations were carried out under vastly diverse weather conditions
because of the relaxed constraints of the program, allowing seeing of
up to 2\arcsec\ and thin cloud coverage.

\begin{figure}[bt]
\centering
\includegraphics[width=8.8cm]{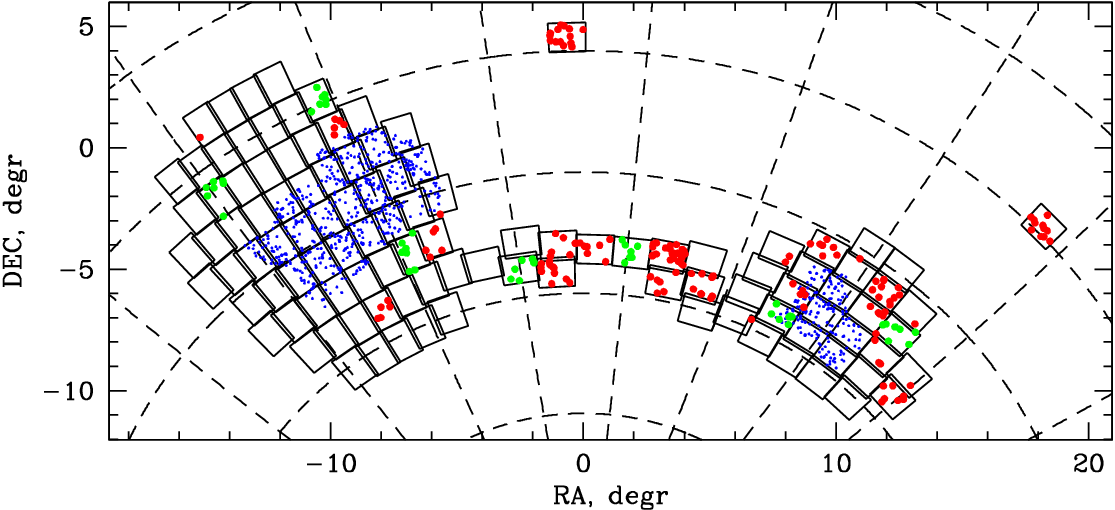} \\
\caption{Location of the spectroscopically followed up objects in 
this work (red) and confirmed QSOs from
\citet[][blue]{2013ApJ...775...92K} and 
\citet[][green]{2016A&A...588A..93I}. The VMC tiles are shown as 
rectangles. The grid shows lines of constant right ascension and 
declination (spaced by 15\,\degr and 5\,\degr, respectively). 
Coordinates are offset with respect to ($\alpha_0$, $\delta_0$) = 
(51\degr, $-$69\degr).}\label{fig:map}
\end{figure}

During the selection we excluded any previously known QSOs and 
gave preference to the brightest candidates in each tile, optimizing
the spectroscopic follow up. Therefore, our results cannot be used 
to draw strict statistical conclusions. Our main goal is to confirm
as many QSOs as possible for future astrometric studies.

Our main target selection was aimed at VLT follow up, but 
spectroscopic confirmation with 
smaller facilities is also possible. To facilitate it, we performed 
an identical candidate selection over the entire VMC survey footprint 
with an additional brightness criterion of $Y$$<$18.0\,mag, to make 
possible a follow up with a 2-m class telescope. This yielded 36 
objects with 15 in the range $Y$=16.5--17.4\,mag. Seven of these were 
observed, bringing the total number of followed up objects to 151.

Table\,\ref{QSO_obs} lists the observed candidates: the VMC 
identification (Col.\,$1$), coordinates (Cols.\,$2$--$3$), magnitudes 
in the $Y$$J$$K_\mathrm{s}$ and their photometric errors 
(Cols.\,$4$--$9$), and the object identification (Col.\,$10$) used in 
the spectroscopic observations. The last consists of the VMC tile 
name and a sequential number in the catalog of sources in that tile; 
the letter g indicates that a source was extended according to the 
VDFS pipeline. \citet{2016A&A...588A..93I} demonstrated that some low 
redshift QSOs fall into that category. The location of the newly 
confirmed QSOs in the ($Y$$-$$J$) versus ($J$$-$$K_\mathrm{s}$) 
color--color diagram is shown in Fig.\,\ref{fig:CCD}, their positions 
on the sky -- in Fig.\,\ref{fig:map} and their $Y$--band finding 
charts -- in Fig.\,\ref{fig:finders}.

\begin{table*}[!htb]
\caption{VMC coordinates and photometry of QSO candidates (in 
order of increasing right ascension). The class indicates if the 
VDFS identified point-like ($-$1) or extended (1) morphology.
For objects in the bright sample for the SAAO 1.9-m telescope we
list (in the last column, in brackets) which tiles they fall in;
more than one tile is listed for objects in overlapping regions.
}\label{QSO_obs}
\begin{center}
\begin{small}
\begin{tabular}{@{}l@{ }r@{ }l@{ }c@{ }c@{ }c@{ }c@{ }c@{ }c@{ }c@{ }p{3.5cm}@{}}
\hline\hline
VMC ID~~~~~~~~~~~~~~~~~~~~~~~~~~~~~~~~~~ & \multicolumn{2}{c}{~~~~~$\alpha$~~~~~~(J2000)~~~~~~$\delta$~~~~~} & $Y$ & $\sigma_Y$ & $J$ & $\sigma_J$ & $K_S$ & $\sigma_{K_S}$ & Class & Object ID \\
 & (h:m:s) & (d:m:s) &~(mag)~&~(mag)~~&~(mag)~&~(mag)~~&~(mag)~&~(mag)~~&  & \\
\hline
VMC J000041.81$-$730258.8 & 00:00:41.81 & $-$73:02:58.8 & 16.868 & 0.007 & 16.334 & 0.006 & 14.986 & 0.006 &    1 & VMC Bright 03 (SMC 4\_1) \\
VMC J000442.82$-$742013.9 & 00:04:42.82 & $-$74:20:13.9 & 18.888 & 0.020 & 18.652 & 0.021 & 17.498 & 0.022 & $-$1 & SMC 3\_1 107 \\
VMC J000505.24$-$741051.5 & 00:05:05.24 & $-$74:10:51.5 & 17.548 & 0.009 & 17.053 & 0.009 & 15.307 & 0.007 &    1 & SMC 3\_1 044g \\
VMC J000611.48$-$645423.7 & 00:06:11.48 & $-$64:54:23.7 & 18.211 & 0.014 & 17.753 & 0.013 & 15.848 & 0.009 &    1 & STR 2\_1 203 \\
VMC J000737.85$-$733445.6 & 00:07:37.85 & $-$73:34:45.6 & 17.878 & 0.011 & 17.729 & 0.013 & 16.398 & 0.012 & $-$1 & SMC 3\_1 119 \\
VMC J000908.49$-$645724.4 & 00:09:08.49 & $-$64:57:24.4 & 18.484 & 0.016 & 18.187 & 0.017 & 17.274 & 0.019 & $-$1 & STR 2\_1 169 \\
VMC J000953.25$-$644109.5 & 00:09:53.25 & $-$64:41:09.5 & 17.884 & 0.012 & 17.701 & 0.013 & 16.489 & 0.013 & $-$1 & STR 2\_1 055 \\
VMC J001001.68$-$650200.2 & 00:10:01.68 & $-$65:02:00.2 & 18.995 & 0.021 & 18.584 & 0.020 & 17.164 & 0.018 &    1 & STR 2\_1 240g \\
VMC J001114.13$-$735110.0 & 00:11:14.13 & $-$73:51:10.0 & 18.234 & 0.014 & 17.717 & 0.012 & 16.357 & 0.012 &    1 & SMC 3\_1 074g \\
VMC J001127.76$-$644304.2 & 00:11:27.76 & $-$64:43:04.2 & 18.622 & 0.017 & 18.186 & 0.016 & 16.727 & 0.014 &    1 & STR 2\_1 225g \\
VMC J001234.62$-$643911.8 & 00:12:34.62 & $-$64:39:11.8 & 18.921 & 0.021 & 18.783 & 0.023 & 17.617 & 0.023 & $-$1 & STR 2\_1 117 \\
VMC J001400.81$-$641325.6 & 00:14:00.81 & $-$64:13:25.6 & 17.800 & 0.011 & 17.792 & 0.013 & 16.217 & 0.011 &    1 & STR 2\_1 174g \\
VMC J001433.39$-$650655.3 & 00:14:33.39 & $-$65:06:55.3 & 18.751 & 0.019 & 18.455 & 0.019 & 16.874 & 0.016 &    1 & STR 2\_1 217g \\
VMC J001530.24$-$644850.8 & 00:15:30.24 & $-$64:48:50.8 & 19.172 & 0.024 & 19.038 & 0.027 & 17.852 & 0.027 & $-$1 & STR 2\_1 100 \\
VMC J001642.69$-$644453.8 & 00:16:42.69 & $-$64:44:53.8 & 19.708 & 0.034 & 19.515 & 0.037 & 18.192 & 0.033 & $-$1 & STR 2\_1 112 \\
VMC J001809.45$-$644458.9 & 00:18:09.45 & $-$64:44:58.9 & 18.494 & 0.016 & 18.538 & 0.020 & 16.992 & 0.017 &    1 & STR 2\_1 077g \\
VMC J002151.98$-$731723.3 & 00:21:51.98 & $-$73:17:23.3 & 18.424 & 0.016 & 17.907 & 0.014 & 16.539 & 0.013 &    1 & SMC 4\_2 014g \\
VMC J002303.23$-$731801.6 & 00:23:03.23 & $-$73:18:01.6 & 19.180 & 0.026 & 18.893 & 0.027 & 17.367 & 0.021 & $-$1 & SMC 4\_2 015 \\
VMC J002322.81$-$710941.4 & 00:23:22.81 & $-$71:09:41.4 & 17.246 & 0.009 & 16.773 & 0.008 & 15.011 & 0.006 &    1 & VMC Bright 08 (SMC 6\_2) \\
VMC J003319.36$-$724305.7 & 00:33:19.36 & $-$72:43:05.7 & 19.664 & 0.038 & 19.354 & 0.037 & 17.846 & 0.029 &    1 & SMC 4\_2 103 \\
VMC J003518.92$-$722855.3 & 00:35:18.92 & $-$72:28:55.3 & 17.627 & 0.010 & 17.377 & 0.011 & 16.335 & 0.012 & $-$1 & SMC 5\_3 128 \\
VMC J003841.18$-$714425.5 & 00:38:41.18 & $-$71:44:25.5 & 19.807 & 0.043 & 19.432 & 0.045 & 18.250 & 0.038 & $-$1 & SMC 5\_3 089 \\
VMC J003856.59$-$703936.4 & 00:38:56.59 & $-$70:39:36.4 & 19.770 & 0.053 & 19.415 & 0.037 & 17.836 & 0.029 & $-$1 & SMC 6\_3 199 \\
VMC J003910.78$-$713409.9 & 00:39:10.78 & $-$71:34:09.9 & 16.977 & 0.007 & 16.483 & 0.007 & 14.537 & 0.005 & $-$1 & SMC 5\_3 021 \\
VMC J003927.58$-$711240.5 & 00:39:27.58 & $-$71:12:40.5 & 19.804 & 0.055 & 19.339 & 0.035 & 18.053 & 0.032 & $-$1 & SMC 6\_3 229 \\
VMC J004006.53$-$705318.5 & 00:40:06.53 & $-$70:53:18.5 & 18.290 & 0.021 & 17.689 & 0.012 & 16.041 & 0.010 &    1 & SMC 6\_3 186g \\
VMC J004037.57$-$710658.2 & 00:40:37.57 & $-$71:06:58.2 & 20.044 & 0.065 & 19.703 & 0.045 & 17.930 & 0.030 & $-$1 & SMC 6\_3 322 \\
VMC J004043.60$-$714541.5 & 00:40:43.60 & $-$71:45:41.5 & 19.444 & 0.032 & 19.125 & 0.036 & 17.609 & 0.025 & $-$1 & SMC 5\_3 075 \\
VMC J004135.79$-$714522.6 & 00:41:35.79 & $-$71:45:22.6 & 18.715 & 0.019 & 18.572 & 0.024 & 17.536 & 0.024 & $-$1 & SMC 5\_3 015 \\
VMC J004144.30$-$702706.2 & 00:41:44.30 & $-$70:27:06.2 & 18.779 & 0.028 & 18.652 & 0.022 & 17.169 & 0.019 &    1 & SMC 6\_3 092g \\
VMC J004625.33$-$711425.7 & 00:46:25.33 & $-$71:14:25.7 & 20.080 & 0.065 & 19.658 & 0.044 & 18.061 & 0.033 & $-$1 & SMC 6\_3 161 \\
VMC J004655.15$-$714451.0 & 00:46:55.15 & $-$71:44:51.0 & 19.287 & 0.029 & 18.910 & 0.030 & 17.240 & 0.020 &    1 & SMC 5\_3 003g \\
VMC J004837.07$-$704955.0 & 00:48:37.07 & $-$70:49:55.0 & 19.767 & 0.053 & 19.154 & 0.030 & 18.285 & 0.037 & $-$1 & SMC 6\_3 195 \\
VMC J004936.13$-$703451.3 & 00:49:36.13 & $-$70:34:51.3 & 18.118 & 0.019 & 17.513 & 0.011 & 16.005 & 0.010 &    1 & SMC 6\_3 141\_2 \\
VMC J004952.18$-$703231.6 & 00:49:52.18 & $-$70:32:31.6 & 19.612 & 0.048 & 19.362 & 0.035 & 17.850 & 0.028 & $-$1 & SMC 6\_3 141 \\
VMC J005116.88$-$710857.1 & 00:51:16.88 & $-$71:08:57.1 & 18.388 & 0.022 & 18.031 & 0.015 & 16.462 & 0.013 & $-$1 & SMC 6\_3 308 \\
VMC J005231.69$-$704705.5 & 00:52:31.69 & $-$70:47:05.5 & 18.160 & 0.020 & 17.927 & 0.014 & 17.053 & 0.018 & $-$1 & SMC 6\_3 310 \\
VMC J010522.22$-$702251.1 & 01:05:22.22 & $-$70:22:51.1 & 16.984 & 0.007 & 17.025 & 0.009 & 15.369 & 0.008 &    1 & VMC Bright 11 (SMC 6\_4, SMC 7\_4) \\
VMC J011548.36$-$704338.0 & 01:15:48.36 & $-$70:43:38.0 & 18.622 & 0.017 & 18.462 & 0.022 & 17.376 & 0.020 & $-$1 & SMC 6\_5 080 \\
VMC J011628.12$-$731447.1 & 01:16:28.12 & $-$73:14:47.1 & 20.122 & 0.046 & 20.034 & 0.060 & 18.514 & 0.041 & $-$1 & SMC 4\_5 060 \\
VMC J011909.15$-$703106.8 & 01:19:09.15 & $-$70:31:06.8 & 18.976 & 0.020 & 18.456 & 0.021 & 16.952 & 0.016 &    1 & SMC 6\_5 086g \\
VMC J011936.67$-$724616.5 & 01:19:36.67 & $-$72:46:16.5 & 20.029 & 0.044 & 19.763 & 0.049 & 18.825 & 0.050 & $-$1 & SMC 4\_5 062 \\
VMC J012143.71$-$710205.5 & 01:21:43.71 & $-$71:02:05.5 & 18.958 & 0.020 & 18.431 & 0.021 & 16.896 & 0.016 &    1 & SMC 6\_5 012g \\
VMC J012146.62$-$723949.1 & 01:21:46.62 & $-$72:39:49.1 & 19.568 & 0.031 & 19.344 & 0.035 & 18.376 & 0.038 & $-$1 & SMC 4\_5 069 \\
VMC J012150.07$-$725023.4 & 01:21:50.07 & $-$72:50:23.4 & 18.627 & 0.017 & 18.300 & 0.017 & 16.385 & 0.012 &    1 & SMC 4\_2 071g \\
VMC J012356.63$-$702107.8 & 01:23:56.63 & $-$70:21:07.8 & 17.934 & 0.011 & 17.588 & 0.013 & 16.494 & 0.013 & $-$1 & SMC 6\_5 114 \\
VMC J012419.21$-$703931.8 & 01:24:19.21 & $-$70:39:31.8 & 19.529 & 0.029 & 19.494 & 0.043 & 17.897 & 0.027 & $-$1 & SMC 6\_5 077 \\
VMC J012545.82$-$703925.8 & 01:25:45.82 & $-$70:39:25.8 & 17.405 & 0.009 & 17.096 & 0.010 & 15.401 & 0.007 &    1 & SMC 6\_5 008g \\
VMC J012808.10$-$723547.1 & 01:28:08.10 & $-$72:35:47.1 & 18.751 & 0.019 & 18.504 & 0.020 & 17.324 & 0.020 & $-$1 & SMC 4\_5 038 \\
VMC J012958.83$-$704859.8 & 01:29:58.83 & $-$70:48:59.8 & 19.104 & 0.022 & 18.897 & 0.029 & 17.873 & 0.027 & $-$1 & SMC 6\_5 052 \\
VMC J013634.37$-$713838.7 & 01:36:34.37 & $-$71:38:38.7 & 17.411 & 0.009 & 17.239 & 0.010 & 15.187 & 0.007 &    1 & SMC 5\_6 047g \\
VMC J013658.86$-$715542.3 & 01:36:58.86 & $-$71:55:42.3 & 18.426 & 0.016 & 18.311 & 0.017 & 16.869 & 0.016 &    1 & SMC 5\_6 057g \\
VMC J014003.42$-$743845.9 & 01:40:03.42 & $-$74:38:45.9 & 17.129 & 0.008 & 16.976 & 0.009 & 15.724 & 0.009 & $-$1 & VMC Bright 24 (SMC 2\_5, BRI 1\_2) \\
VMC J020706.19$-$741848.2 & 02:07:06.19 & $-$74:18:48.2 & 18.514 & 0.019 & 18.434 & 0.020 & 16.988 & 0.017 &    1 & BRI 2\_3 195g \\
VMC J020721.28$-$742456.6 & 02:07:21.28 & $-$74:24:56.6 & 17.874 & 0.012 & 17.706 & 0.014 & 16.445 & 0.013 & $-$1 & BRI 2\_3 007 \\
VMC J021022.07$-$733017.1 & 02:10:22.07 & $-$73:30:17.1 & 18.730 & 0.022 & 18.353 & 0.020 & 17.147 & 0.019 & $-$1 & BRI 2\_3 011 \\
VMC J021044.79$-$733423.0 & 02:10:44.79 & $-$73:34:23.0 & 18.414 & 0.018 & 18.166 & 0.017 & 17.020 & 0.017 & $-$1 & BRI 2\_3 143 \\
VMC J021335.30$-$742109.1 & 02:13:35.30 & $-$74:21:09.1 & 19.169 & 0.028 & 18.816 & 0.026 & 17.263 & 0.020 &    1 & BRI 2\_3 238g \\
VMC J021353.93$-$742048.2 & 02:13:53.93 & $-$74:20:48.2 & 18.235 & 0.016 & 17.892 & 0.015 & 16.891 & 0.016 & $-$1 & BRI 2\_3 159 \\
VMC J021644.16$-$733228.4 & 02:16:44.16 & $-$73:32:28.4 & 20.532 & 0.081 & 20.344 & 0.078 & 19.506 & 0.086 & $-$1 & BRI 2\_3 032 \\
\hline 
\end{tabular}
\end{small}
\end{center}
\end{table*}
\addtocounter{table}{-1}

\begin{table*}[!htb]
\caption{Continued.}
\begin{center}
\begin{small}
\begin{tabular}{@{}l@{ }r@{ }l@{ }c@{ }c@{ }c@{ }c@{ }c@{ }c@{ }c@{ }l@{}}
\hline\hline
VMC ID~~~~~~~~~~~~~~~~~~~~~~~~~~~~~~~~~~ & \multicolumn{2}{c}{~~~~~$\alpha$~~~~~~(J2000)~~~~~~$\delta$~~~~~} & $Y$ & $\sigma_Y$ & $J$ & $\sigma_J$ & $K_S$ & $\sigma_{K_S}$ & Class & Object ID \\
 & (h:m:s) & (d:m:s) &~(mag)~&~(mag)~~&~(mag)~&~(mag)~~&~(mag)~&~(mag)~~&  & \\
\hline
VMC J021905.21$-$741445.5 & 02:19:05.21 & $-$74:14:45.5 & 18.336 & 0.017 & 18.214 & 0.018 & 16.894 & 0.016 & $-$1 & BRI 2\_3 069 \\
VMC J022232.59$-$733827.9 & 02:22:32.59 & $-$73:38:27.9 & 18.808 & 0.044 & 18.592 & 0.054 & 17.853 & 0.037 & $-$1 & BRI 2\_3 057 \\
VMC J022822.54$-$732303.1 & 02:28:22.54 & $-$73:23:03.1 & 18.118 & 0.012 & 18.085 & 0.016 & 16.570 & 0.014 &    1 & BRI 3\_4 117 \\
VMC J022842.84$-$725447.9 & 02:28:42.84 & $-$72:54:47.9 & 18.813 & 0.019 & 18.671 & 0.023 & 17.481 & 0.022 & $-$1 & BRI 3\_4 141 \\
VMC J022923.80$-$731952.1 & 02:29:23.80 & $-$73:19:52.1 & 18.573 & 0.016 & 18.088 & 0.016 & 16.468 & 0.013 &    1 & BRI 3\_4 109g \\
VMC J022941.30$-$724805.0 & 02:29:41.30 & $-$72:48:05.0 & 18.974 & 0.021 & 18.579 & 0.022 & 17.381 & 0.021 &    1 & BRI 3\_4 209g \\
VMC J022955.38$-$730447.6 & 02:29:55.38 & $-$73:04:47.6 & 18.849 & 0.019 & 18.496 & 0.020 & 17.390 & 0.021 & $-$1 & BRI 3\_4 125 \\
VMC J023009.38$-$730346.1 & 02:30:09.38 & $-$73:03:46.1 & 17.922 & 0.011 & 17.632 & 0.012 & 16.659 & 0.014 & $-$1 & BRI 3\_4 131 \\
VMC J023043.89$-$724536.2 & 02:30:43.89 & $-$72:45:36.2 & 19.415 & 0.028 & 19.284 & 0.034 & 18.173 & 0.035 & $-$1 & BRI 3\_4 129 \\
VMC J023108.46$-$734439.3 & 02:31:08.46 & $-$73:44:39.3 & 20.091 & 0.045 & 19.737 & 0.043 & 18.291 & 0.037 & $-$1 & BRI 2\_4 054 \\
VMC J023121.22$-$731654.4 & 02:31:21.22 & $-$73:16:54.4 & 20.547 & 0.062 & 20.101 & 0.062 & 18.470 & 0.042 & $-$1 & BRI 3\_4 089 \\
VMC J023243.92$-$725711.6 & 02:32:43.92 & $-$72:57:11.6 & 19.542 & 0.030 & 19.441 & 0.039 & 18.165 & 0.034 & $-$1 & BRI 3\_4 046 \\
VMC J023225.48$-$725915.6 & 02:32:25.48 & $-$72:59:15.6 & 18.388 & 0.015 & 18.129 & 0.016 & 17.635 & 0.025 & $-$1 & BRI 3\_4 046\_2 \\
VMC J023421.27$-$731116.5 & 02:34:21.27 & $-$73:11:16.5 & 18.978 & 0.021 & 18.503 & 0.020 & 16.922 & 0.016 &    1 & BRI 3\_4 041g \\
VMC J023422.51$-$723648.4 & 02:34:22.51 & $-$72:36:48.4 & 19.162 & 0.023 & 18.776 & 0.025 & 17.143 & 0.018 &    1 & BRI 3\_4 190g \\
VMC J023448.42$-$725828.5 & 02:34:48.42 & $-$72:58:28.5 & 17.407 & 0.009 & 17.099 & 0.009 & 15.337 & 0.007 & $-$1 & BRI 3\_4 153 \\
VMC J023528.23$-$730158.2 & 02:35:28.23 & $-$73:01:58.2 & 17.910 & 0.011 & 17.897 & 0.014 & 16.450 & 0.013 & $-$1 & BRI 3\_4 157 \\
VMC J023539.49$-$743705.2 & 02:35:39.49 & $-$74:37:05.2 & 19.016 & 0.021 & 18.899 & 0.024 & 17.535 & 0.023 & $-$1 & BRI 2\_4 025 \\
VMC J023550.79$-$731710.4 & 02:35:50.79 & $-$73:17:10.4 & 20.043 & 0.043 & 19.594 & 0.042 & 18.067 & 0.032 & $-$1 & BRI 3\_4 166 \\
VMC J023716.56$-$725030.7 & 02:37:16.56 & $-$72:50:30.7 & 18.661 & 0.017 & 18.495 & 0.020 & 17.112 & 0.018 & $-$1 & BRI 3\_4 003 \\
VMC J023728.54$-$724242.6 & 02:37:28.54 & $-$72:42:42.6 & 20.328 & 0.054 & 20.197 & 0.068 & 18.978 & 0.059 & $-$1 & BRI 3\_4 004 \\
VMC J023904.89$-$741504.2 & 02:39:04.89 & $-$74:15:04.2 & 18.723 & 0.018 & 18.596 & 0.020 & 17.460 & 0.022 & $-$1 & BRI 2\_4 121 \\
VMC J023919.57$-$744321.5 & 02:39:19.57 & $-$74:43:21.5 & 19.248 & 0.024 & 18.996 & 0.026 & 17.370 & 0.021 &    1 & BRI 2\_4 014g \\
VMC J023937.56$-$724942.3 & 02:39:37.56 & $-$72:49:42.3 & 19.328 & 0.026 & 19.258 & 0.034 & 18.049 & 0.031 & $-$1 & BRI 3\_4 026 \\
VMC J024049.48$-$741133.4 & 02:40:49.48 & $-$74:11:33.4 & 20.794 & 0.077 & 20.504 & 0.078 & 19.220 & 0.070 & $-$1 & BRI 2\_4 150 \\
VMC J024401.00$-$730831.2 & 02:44:01.00 & $-$73:08:31.2 & 19.342 & 0.026 & 18.922 & 0.027 & 17.414 & 0.022 &    1 & BRI 3\_4 437g \\
VMC J024458.46$-$740624.4 & 02:44:58.46 & $-$74:06:24.4 & 20.675 & 0.072 & 20.447 & 0.076 & 19.677 & 0.101 & $-$1 & BRI 2\_4 014 \\
VMC J024501.04$-$724116.8 & 02:45:01.04 & $-$72:41:16.8 & 18.672 & 0.018 & 18.289 & 0.018 & 16.623 & 0.014 &    1 & BRI 3\_4 218g \\
VMC J024541.60$-$731211.8 & 02:45:41.60 & $-$73:12:11.8 & 18.064 & 0.012 & 17.604 & 0.012 & 15.948 & 0.010 &    1 & BRI 3\_4 347g \\
VMC J024555.70$-$731435.5 & 02:45:55.70 & $-$73:14:35.5 & 19.140 & 0.023 & 19.097 & 0.030 & 17.484 & 0.022 &    1 & BRI 3\_4 271g \\
VMC J024634.70$-$724504.3 & 02:46:34.70 & $-$72:45:04.3 & 19.165 & 0.024 & 18.835 & 0.026 & 17.286 & 0.020 &    1 & BRI 3\_4 050g \\
VMC J024635.48$-$725142.5 & 02:46:35.48 & $-$72:51:42.5 & 18.311 & 0.014 & 18.384 & 0.019 & 17.151 & 0.019 & $-$1 & BRI 3\_4 033 \\
VMC J030948.59$-$724423.3 & 03:09:48.59 & $-$72:44:23.3 & 18.884 & 0.022 & 18.681 & 0.021 & 17.806 & 0.027 & $-$1 & BRI 3\_6 025 \\
VMC J031128.66$-$734010.5 & 03:11:28.66 & $-$73:40:10.5 & 19.487 & 0.032 & 19.113 & 0.029 & 17.481 & 0.023 & $-$1 & BRI 3\_6 031 \\
VMC J031509.77$-$730025.0 & 03:15:09.77 & $-$73:00:25.0 & 18.163 & 0.014 & 17.963 & 0.014 & 16.579 & 0.014 & $-$1 & BRI 3\_6 068 \\
VMC J032143.49$-$732110.9 & 03:21:43.49 & $-$73:21:10.9 & 17.415 & 0.009 & 17.107 & 0.009 & 15.294 & 0.007 & $-$1 & BRI 3\_6 143 \\
VMC J032158.17$-$730440.4 & 03:21:58.17 & $-$73:04:40.4 & 18.490 & 0.017 & 18.396 & 0.018 & 17.331 & 0.020 & $-$1 & BRI 3\_6 145 \\
VMC J032328.65$-$730310.9 & 03:23:28.65 & $-$73:03:10.9 & 20.128 & 0.053 & 19.559 & 0.038 & 17.808 & 0.027 & $-$1 & BRI 3\_6 090 \\
VMC J032404.61$-$640722.3 & 03:24:04.61 & $-$64:07:22.3 & 18.241 & 0.014 & 18.112 & 0.017 & 16.752 & 0.015 &    1 & STR 1\_1 226g \\
VMC J032550.80$-$725506.3 & 03:25:50.80 & $-$72:55:06.3 & 19.546 & 0.034 & 19.281 & 0.032 & 18.074 & 0.032 & $-$1 & BRI 3\_6 100 \\
VMC J032812.35$-$645019.1 & 03:28:12.35 & $-$64:50:19.1 & 18.853 & 0.020 & 18.492 & 0.021 & 17.484 & 0.022 &    1 & STR 1\_1 033g \\
VMC J032812.54$-$725541.5 & 03:28:12.54 & $-$72:55:41.5 & 19.090 & 0.024 & 19.242 & 0.040 & 17.821 & 0.028 & $-$1 & BRI 3\_7 043 \\
VMC J032818.72$-$725633.6 & 03:28:18.72 & $-$72:56:33.6 & 18.434 & 0.016 & 17.990 & 0.018 & 16.331 & 0.012 &    1 & BRI 3\_7 055g \\
VMC J032832.83$-$644055.8 & 03:28:32.83 & $-$64:40:55.8 & 18.872 & 0.021 & 18.550 & 0.022 & 17.388 & 0.021 & $-$1 & STR 1\_1 106 \\
VMC J032850.47$-$642418.4 & 03:28:50.47 & $-$64:24:18.4 & 19.939 & 0.042 & 19.619 & 0.045 & 18.102 & 0.033 & $-$1 & STR 1\_1 108 \\
VMC J032903.37$-$640515.7 & 03:29:03.37 & $-$64:05:15.7 & 19.285 & 0.027 & 19.032 & 0.030 & 17.612 & 0.025 & $-$1 & STR 1\_1 034 \\
VMC J033054.45$-$635727.3 & 03:30:54.45 & $-$63:57:27.3 & 18.880 & 0.021 & 18.404 & 0.019 & 16.793 & 0.015 &    1 & STR 1\_1 097g \\
VMC J033121.11$-$644637.6 & 03:31:21.11 & $-$64:46:37.6 & 19.132 & 0.024 & 19.082 & 0.029 & 17.897 & 0.028 & $-$1 & STR 1\_1 128 \\
VMC J033126.71$-$732415.3 & 03:31:26.71 & $-$73:24:15.3 & 20.131 & 0.049 & 19.830 & 0.063 & 18.106 & 0.033 & $-$1 & BRI 3\_7 127 \\
VMC J033206.62$-$743318.7 & 03:32:06.62 & $-$74:33:18.7 & 20.603 & 0.065 & 20.275 & 0.070 & 18.902 & 0.059 & $-$1 & BRI 2\_7 089 \\
VMC J033207.40$-$635514.5 & 03:32:07.40 & $-$63:55:14.5 & 18.288 & 0.015 & 18.088 & 0.016 & 16.568 & 0.014 &    1 & STR 1\_1 100g \\
VMC J033227.01$-$643655.8 & 03:32:27.01 & $-$64:36:55.8 & 19.426 & 0.029 & 19.210 & 0.034 & 17.733 & 0.026 & $-$1 & STR 1\_1 083 \\
VMC J033315.61$-$640607.8 & 03:33:15.61 & $-$64:06:07.8 & 18.783 & 0.020 & 18.336 & 0.019 & 16.741 & 0.015 &    1 & STR 1\_1 072g \\
VMC J033405.15$-$742400.3 & 03:34:05.15 & $-$74:24:00.3 & 19.983 & 0.040 & 19.921 & 0.052 & 18.590 & 0.046 & $-$1 & BRI 2\_7 130 \\
VMC J033412.54$-$643623.8 & 03:34:12.54 & $-$64:36:23.8 & 16.577 & 0.006 & 16.267 & 0.006 & 15.025 & 0.006 & $-$1 & VMC Bright 05 (STR 2\_1) \\
VMC J033438.94$-$724116.9 & 03:34:38.94 & $-$72:41:16.9 & 18.765 & 0.020 & 18.496 & 0.024 & 17.237 & 0.019 & $-$1 & BRI 3\_7 182 \\
VMC J033603.89$-$641431.8 & 03:36:03.89 & $-$64:14:31.8 & 18.577 & 0.017 & 18.206 & 0.018 & 17.151 & 0.019 & $-$1 & STR 1\_1 046 \\
VMC J033611.01$-$642017.6 & 03:36:11.01 & $-$64:20:17.6 & 19.659 & 0.035 & 19.421 & 0.040 & 18.241 & 0.036 & $-$1 & STR 1\_1 048 \\
VMC J033613.21$-$643309.8 & 03:36:13.21 & $-$64:33:09.8 & 19.480 & 0.031 & 19.137 & 0.032 & 18.100 & 0.033 & $-$1 & STR 1\_1 047 \\
VMC J033836.66$-$723001.7 & 03:38:36.66 & $-$72:30:01.7 & 18.036 & 0.013 & 17.619 & 0.014 & 15.935 & 0.010 &    1 & BRI 3\_7 154g \\
VMC J034040.77$-$740750.1 & 03:40:40.77 & $-$74:07:50.1 & 18.998 & 0.021 & 18.887 & 0.025 & 17.691 & 0.025 & $-$1 & BRI 2\_7 145 \\
VMC J034052.05$-$735436.6 & 03:40:52.05 & $-$73:54:36.6 & 19.100 & 0.023 & 18.733 & 0.023 & 17.417 & 0.022 & $-$1 & BRI 2\_7 018 \\
VMC J034123.04$-$735156.7 & 03:41:23.04 & $-$73:51:56.7 & 17.538 & 0.009 & 17.062 & 0.009 & 15.745 & 0.009 &    1 & BRI 2\_7 053g \\
VMC J034129.52$-$731500.8 & 03:41:29.52 & $-$73:15:00.8 & 17.979 & 0.013 & 17.484 & 0.013 & 15.838 & 0.009 &    1 & BRI 3\_7 176g \\
\hline 
\end{tabular}
\end{small}
\end{center}
\end{table*}
\addtocounter{table}{-1}

\begin{table*}[!htb]
\caption{Continued.}
\begin{center}
\begin{small}
\begin{tabular}{@{}l@{ }r@{ }l@{ }c@{ }c@{ }c@{ }c@{ }c@{ }c@{ }c@{ }l@{}}
\hline\hline
VMC ID~~~~~~~~~~~~~~~~~~~~~~~~~~~~~~~~~~ & \multicolumn{2}{c}{~~~~~$\alpha$~~~~~~(J2000)~~~~~~$\delta$~~~~~} & $Y$ & $\sigma_Y$ & $J$ & $\sigma_J$ & $K_S$ & $\sigma_{K_S}$ & Class & Object ID \\
 & (h:m:s) & (d:m:s) &~(mag)~&~(mag)~~&~(mag)~&~(mag)~~&~(mag)~&~(mag)~~&  & \\
\hline
VMC J034215.07$-$732514.7 & 03:42:15.07 & $-$73:25:14.7 & 19.591 & 0.033 & 19.455 & 0.047 & 18.393 & 0.039 & $-$1 & BRI 3\_7 019 \\
VMC J034250.73$-$743335.9 & 03:42:50.73 & $-$74:33:35.9 & 19.856 & 0.037 & 19.383 & 0.035 & 17.560 & 0.023 & $-$1 & BRI 2\_7 021 \\
VMC J034257.34$-$734609.6 & 03:42:57.34 & $-$73:46:09.6 & 19.117 & 0.022 & 18.925 & 0.025 & 17.335 & 0.021 & $-$1 & BRI 2\_7 060 \\
VMC J034302.85$-$733814.7 & 03:43:02.85 & $-$73:38:14.7 & 19.366 & 0.027 & 19.268 & 0.032 & 17.880 & 0.029 & $-$1 & BRI 2\_7 064 \\
VMC J034743.46$-$734119.0 & 03:47:43.46 & $-$73:41:19.0 & 19.263 & 0.025 & 19.042 & 0.028 & 18.067 & 0.032 & $-$1 & BRI 2\_7 158 \\
VMC J034806.80$-$740039.4 & 03:48:06.80 & $-$74:00:39.4 & 18.525 & 0.016 & 18.202 & 0.016 & 16.921 & 0.016 & $-$1 & BRI 2\_7 151 \\
VMC J034817.21$-$735138.3 & 03:48:17.21 & $-$73:51:38.3 & 19.182 & 0.024 & 18.873 & 0.025 & 17.825 & 0.028 &    1 & BRI 2\_7 042g \\
VMC J043425.74$-$705453.8 & 04:34:25.74 & $-$70:54:53.8 & 17.107 & 0.008 & 16.751 & 0.008 & 15.162 & 0.007 &    1 & VMC Bright 14 (LMC 5\_1) \\
VMC J043854.34$-$712613.3 & 04:38:54.34 & $-$71:26:13.3 & 18.220 & 0.013 & 17.858 & 0.017 & 16.797 & 0.016 & $-$1 & LMC 4\_2 066 \\
VMC J043914.87$-$722124.0 & 04:39:14.87 & $-$72:21:24.0 & 18.468 & 0.016 & 18.220 & 0.021 & 16.941 & 0.017 & $-$1 & LMC 4\_2 038 \\
VMC J044015.31$-$713128.8 & 04:40:15.31 & $-$71:31:28.8 & 20.144 & 0.049 & 19.870 & 0.071 & 18.911 & 0.062 & $-$1 & LMC 4\_2 043 \\
VMC J044616.30$-$722827.2 & 04:46:16.30 & $-$72:28:27.2 & 17.881 & 0.011 & 17.787 & 0.016 & 16.681 & 0.015 & $-$1 & LMC 4\_2 120 \\
VMC J044713.31$-$720908.1 & 04:47:13.31 & $-$72:09:08.1 & 18.962 & 0.021 & 18.838 & 0.032 & 17.765 & 0.028 &    1 & LMC 4\_2 033g \\
VMC J050031.54$-$660908.4 & 05:00:31.54 & $-$66:09:08.4 & 20.447 & 0.081 & 20.257 & 0.111 & 19.364 & 0.090 & $-$1 & LMC 8\_3 013 \\
VMC J050132.72$-$655606.2 & 05:01:32.72 & $-$65:56:06.2 & 18.638 & 0.019 & 18.353 & 0.024 & 17.092 & 0.018 & $-$1 & LMC 8\_3 039 \\
VMC J050258.90$-$654815.7 & 05:02:58.90 & $-$65:48:15.7 & 20.546 & 0.089 & 20.197 & 0.105 & 18.859 & 0.062 & $-$1 & LMC 8\_3 023 \\
VMC J050431.38$-$660723.7 & 05:04:31.38 & $-$66:07:23.7 & 19.379 & 0.034 & 19.076 & 0.041 & 17.761 & 0.028 &    1 & LMC 8\_3 095g \\
VMC J050546.58$-$662327.4 & 05:05:46.58 & $-$66:23:27.4 & 20.455 & 0.081 & 20.204 & 0.106 & 19.011 & 0.067 & $-$1 & LMC 8\_3 016 \\
VMC J051713.17$-$732818.9 & 05:17:13.17 & $-$73:28:18.9 & 20.040 & 0.059 & 19.847 & 0.063 & 18.335 & 0.040 & $-$1 & LMC 3\_5 091 \\
VMC J051754.32$-$734354.1 & 05:17:54.32 & $-$73:43:54.1 & 18.795 & 0.024 & 18.488 & 0.023 & 16.652 & 0.015 &    1 & LMC 3\_5 085g \\
VMC J052234.44$-$733523.7 & 05:22:34.44 & $-$73:35:23.7 & 18.247 & 0.017 & 18.125 & 0.018 & 17.073 & 0.018 & $-$1 & LMC 3\_5 078 \\
VMC J052541.50$-$735734.3 & 05:25:41.50 & $-$73:57:34.3 & 18.369 & 0.018 & 18.305 & 0.020 & 16.664 & 0.014 &    1 & LMC 3\_5 063g \\
VMC J052727.83$-$735609.9 & 05:27:27.83 & $-$73:56:09.9 & 20.539 & 0.089 & 20.232 & 0.083 & 18.538 & 0.046 & $-$1 & LMC 3\_5 103 \\
VMC J055137.33$-$635018.9 & 05:51:37.33 & $-$63:50:18.9 & 16.548 & 0.006 & 16.138 & 0.006 & 15.129 & 0.007 &    1 & VMC Bright 02 (LMC 10\_7) \\
VMC J235716.69$-$733833.0 & 23:57:16.69 & $-$73:38:33.0 & 18.681 & 0.018 & 18.357 & 0.018 & 17.306 & 0.020 & $-$1 & SMC 3\_1 020 \\
VMC J235836.89$-$733253.8 & 23:58:36.89 & $-$73:32:53.8 & 19.888 & 0.040 & 19.656 & 0.043 & 18.104 & 0.032 & $-$1 & SMC 3\_1 035 \\
VMC J235848.46$-$734953.6 & 23:58:48.46 & $-$73:49:53.6 & 19.306 & 0.026 & 19.122 & 0.030 & 18.070 & 0.032 & $-$1 & SMC 3\_1 043 \\
\hline 
\end{tabular}
\end{small}
\end{center}
\end{table*}

Appendix\,\ref{app:compl_sample} presents the complete list of
all VMC survey quasar candidates in latest internal release (Aug
2022) that match our color and variability criteria, for the
benefit of further studies. The color criterion alone yields
163226 objects, and 3609 of them have statistically significant
slopes (Slope/$\sigma_{\rm Slope}$$\geq$3) that exceed the
|Slope|=10$^{-4}$\,mag\,day$^{-1}$ limit.

\section{Spectroscopic follow up observations}\label{sec:spectra}

We selected for follow up 142 candidates, among the brightest
in each tile and well scattered across the survey footprint to
improve the chance for observing in service mode. Two additional
objects
(BRI 3\_4 046\_2 and SMC 6\_3 141\_2) serendipitously fell into
the slits, increasing the total number of observed targets to
144. The selection and the observing followed the same procedure
as in \citet{2016A&A...588A..93I}. We used FORS2 \citep[FOcal
Reducer and low dispersion Spectrograph;][]{1998Msngr..94....1A}
at the VLT (Very Large Telescope) between Oct 2016 and
Aug 2017 in long-slit mode with the 300V+10 grism, GG435+81 order 
sorting filter, and 1.3\arcsec\ wide slit. The spectra cover
$\lambda$=445--865\,nm with a resolving power
$R$=$\lambda$/$\Delta\lambda$$\sim$440. 
Two exposures with integration times between 60\,sec and 530\,seconds
were taken, depending on the QSO's apparent brightness at the 
time of the VMC observations, which is not necessarily the same 
as at the moment of the follow up because of the intrinsic QSO 
variability that may render the quasars fainter or brighter.
In some cases we
had more exposures because the conditions were too poor even for 
our relaxed weather constraints or the observations were 
interrupted because the telescopes were closed and a full set of 
observations was obtained on another night. We also exploited
the low quality data as long as emission lines were identifiable.
The signal--to--noise ratio is 
typically $\sim$10--20 at the centre of the wavelength range, at 
continuum level and higher at the emission lines. The reduction 
was performed with the FORS2 pipeline (v.\,5.3.23). The 
spectrophotometric calibration was carried out with standards 
\citep{1990AJ.....99.1621O,1992PASP..104..533H,1994PASP..106..566H,2014Msngr.158...16M, 2014A&A...568A...9M}, 
observed and processed as the science spectra. The VLT observing 
log is given in Table\,\ref{tab:log1}.

An additional seven objects (bringing the total number of spectra
to 151) were observed with  
SpUpNIC\footnote{\url{https://www.saao.ac.za/astronomers/spupnic/}}
\citep[Spectrograph Upgrade: Newly Improved Cassegrain;][]{2016SPIE.9908E..27C,2019JATIS...5b4007C}
long slit spectrograph at the 1.9-m telescope at SAAO in Nov 2017. 
Grating 7 with no order separation filter (to extend the spectral 
coverage; emission lines from the second order spectra were 
ignored) was used with a 1.95\arcsec\ wide slit, delivering a
resolving power of R$\sim$700-1700 over 
$\lambda$$\lambda$=3750--9300\,\AA. The slit was always oriented 
East-West and a single 1800\,s exposure was taken for each object. 
The data reduction was performed in the standard way, with bias 
and dome flat field corrections. The wavelength calibration was 
done with spectra from an internal CuAr lamp, obtained after each 
science observation to cancel out possible instrument flexures.
Feige\,110 and LTT\,3218 standards (from the same lists as for 
the VLT) were used to derive the spectral response. The 1.9-m 
telescope observing log is given in Table\,\ref{tab:log2}.

The reduced 148 spectra (for three objects no 1-dimensional 
spectra could be extracted, usually due to poor weather 
conditions) are shown in Fig.\,\ref{fig:spectra}. The emission 
lines were identified by comparing our spectra with a
composite QSO spectrum \citep{2001AJ....122..549V}. Usually, we 
had multiple lines; if only one was available, it had to be 
Mg{\sc ii} 2798\,\AA, because if the line belonged to a
different element or if the redshift was different, other 
emission lines with comparable strength would have been inside 
the observed wavelength range (given that the S/N was
sufficient to detect the Mg{\sc ii}), and such lines were missing.
The redshifts were measured as in \citet{2016A&A...588A..93I},
fitting emission lines with Gaussian profiles using the 
IRAF\footnote{The Image Reduction and Analysis Facility is 
distributed by the National Optical Astronomy Observatory, which 
is operated by the Association of Universities for Research in 
Astronomy (AURA) under a cooperative agreement with the National 
Science Foundation.} task {\it splot}. The results are listed in 
Table\,\ref{tab:redshifts}. Some features at the edges of the 
spectra or suffering heavy contamination by sky emission lines or
affected by intervening absorption were omitted from the analysis. 
The statistical errors and errors from wavelength calibration are 
negligible \citep[see Sect.\,3 in ][]{2016A&A...588A..93I}. We
measured the redshift error from different lines of the same object. 
The finding charts of followed up objects are shown in
Fig.\,\ref{app:finder_charts}.

A single line was observed in some spectra. For example, this
occurs for quasars at redshifts between $\sim$1 and $\sim$1.4 when
the only prominent line within our wavelength range is Mg{\sc ii}
2798\,\AA. The featureless continuum outside of the line excludes
the presence of any other lines: e.g., if the observed line was
C{\sc iii}] 1909, then C{\sc iv} 1549 should have appeared in the
blue part of the spectrum. In most cases the spectra have
sufficient S/N (we adopt a limit of 10) at the continuum level to
exclude the presence of other lines, but not always; therefore,
some of our redshifts are tentative, and they are marked in
Table\,\ref{tab:redshifts} with colons signs. The table also
contains the S/N calculated in the vicinity of each emission line
that we used for redshift evaluation. The S/N refers to resolution
elements, so for the broad lines the real S/N is much higher and
especially for the redshift estimates the derived errors for the
central wavelength are more relevant. For narrow line objects the
estimated S/N is dominated by the continuum level and is a lower
limit for core of the line (e.g., in LMC 3\_5 085g,
Bridge 3\_4 209g, Bridge 2\_7 053g). Finally, for the single line
objects we tentatively assign redshift errors of 0.005 for
$z$$\leq$1 and 0.015 for $z$$>$1 -- typical values at these
redshifts.

\begin{table}[!htb]
\caption{Log of the VLT spectroscopic observations. Starting times, 
exposure times, starting and ending airmasses, and slit position 
angles for each exposure are listed on separate successive lines. 
Multiple exposures of each object are entered separately. Only the 
observations of the first three objects are shown for guidance, the 
entire table is included on the electronic edition.}\label{tab:log1}
\begin{center}
\begin{small}
\begin{tabular}{@{}c@{ }c@{ }c@{ }c@{ }r@{}}
\hline\hline
Object  ID     & UT at start of obs.     & Exp. & sec\,$z$     &~~~Slit\,PA~~\\
               & ~~yyyy-mm-ddThh:mm:ss~~ & (s)  & (dex)        & (deg)     \\
\hline
BRI\,2\_3\,007 & ~2016-12-08T05:42:13.371 &  60 & 1.912$-$1.916 & $-$65.199 \\
BRI\,2\_3\,007 & ~2016-12-08T05:43:58.280 &  60 & 1.920$-$1.923 & $-$65.199 \\
BRI\,2\_3\,011 & ~2016-12-05T00:26:16.593 & 180 & 1.562$-$1.558 &    30.072 \\
BRI\,2\_3\,011 & ~2016-12-05T00:30:04.451 & 180 & 1.559$-$1.556 &    30.072 \\
BRI\,2\_3\,032 & ~2016-12-07T06:00:32.486 & 510 & 1.913$-$1.945 & $-$67.766 \\ 
BRI\,2\_3\,032 & ~2016-12-07T06:10:01.053 & 510 & 1.950$-$1.984 & $-$67.766 \\ 
\hline
\end{tabular}
\end{small}
\end{center}
\end{table}

\begin{table}[!htb]
\caption{Log of the 1.9-m SAAO spectroscopic observations. Starting 
times and the airmasses at the start of each exposure are listed. A
single 1800\,sec exposure was obtained for each object.}\label{tab:log2}
\begin{center}
\begin{small}
\begin{tabular}{@{}c@{ }c@{ }c@{}}
\hline\hline
Object  ID      & UT at start of obs.   & sec\,$z$ \\
                &~~yyyy-mm-ddThh:mm:ss~~& (dex)    \\
\hline
VMC SAAO 01 I02 & 2017-11-23T00:15:11 & 1.17     \\
VMC SAAO 02 I03 & 2017-11-27T19:41:13 & 1.35     \\
VMC SAAO 04 I05 & 2017-11-22T23:23:11 & 1.21     \\
VMC SAAO 11 I08 & 2017-11-23T20:01:21 & 1.31     \\
VMC SAAO 08 I11 & 2017-11-24T19:55:06 & 1.27     \\
VMC SAAO 10 I14 & 2017-11-28T22:38:57 & 1.28     \\
VMC SAAO 15 I24 & 2017-11-28T21:54:58 & 1.42     \\
\hline
\end{tabular}
\end{small}
\end{center}
\end{table}

\begin{table}[]
\caption{Derived parameters for the objects in this paper: detected
spectral features and their central wavelengths, signal to noise
ratios within $\pm$20\AA\ from the line centers measured with DER\_SNR
\citep{2008ASPC..394..505S}, estimated redshifts (colon marks redshifts
we consider tentative due to single line identification and low S/N;
a second redshift from Quaia is listed for 74 objects in common) and
object classifications.}\label{tab:redshifts}
\begin{center}
\begin{small}
\begin{tabular}{@{}l@{}c@{}c@{}c@{}c@{}}
\hline\hline
Object  ID       & Spectral feature, observed & ~S/N~  &~~~Redshift~~~~~~       & Class.\\
                 & central wavelength (\AA)   &        & $z$                    & \\
\hline
BRI 2\_4 054     & O{\sc i}/Si{\sc ii}\,1305: 5072.93$\pm$2.27 & 14 & 2.881$\pm$0.004   & QSO \\
                 & C{\sc iv} 1549:            6012.24$\pm$1.35 &  3 &                   & \\
                 & C{\sc iii}] 1909:          7400.55$\pm$1.90 &  3 &                   & \\
BRI 3\_4 131     & O{\sc i}/Si{\sc ii}\,1305: 5008.19$\pm$0.42 &  5 & 2.838$\pm$0.001   & QSO \\
                 & C{\sc iv} 1549:            5946.14$\pm$0.45 & 14 & (2.82$\pm$0.10)   & \\
                 & C{\sc iii}] 1909:          7326.39$\pm$1.84 & 10 &                   & \\
BRI 3\_6 090     & C{\sc iv} 1549:            5807.74$\pm$2.46 &  4 & 2.748$\pm$0.002   & QSO \\
                 & C{\sc iii}] 1909:          7151.63$\pm$1.51 & 15 &                   & \\
BRI 2\_7 158     & C{\sc iv} 1549:            5815.70$\pm$1.42 & 28 & 2.746$\pm$0.017   & QSO \\
                 & C{\sc iii}] 1909:          7134.48$\pm$3.29 & 20 & (2.78$\pm$0.25)   & \\
BRI 3\_6 025     & Si{\sc ii} 1263:           4649.72$\pm$3.25 & 15 & 2.66$\pm$0.02     & QSO \\
                 & O{\sc i}/Si{\sc ii}\,1305: 4799.15$\pm$2.37 & 13 & (2.65$\pm$0.36)   & \\
                 & C{\sc ii} 1335:            4916.89$\pm$2.55 & 10 &                   & \\
                 & Si{\sc iv}/O{\sc iv}\,1397:5119.43$\pm$1.62 & 24 &                   & \\
                 & C{\sc iv} 1549:            5642.04$\pm$2.89 & 29 &                   & \\
                 & C{\sc iii}] 1909:          6939.06$\pm$8.70 &  6 &                   & \\
SMC 4\_5 069     & Si{\sc iv}/O{\sc iv}\,1397:5105.90$\pm$2.32 & 19 & 2.652$\pm$0.004   & QSO \\
                 & C{\sc iv} 1549:            5649.85$\pm$0.87 & 13 & (2.65$\pm$0.10)   & \\
                 & C{\sc iii}] 1909:          6974.87$\pm$0.86 & 12 &                   & \\
SMC 6\_3 310     & C{\sc iv} 1549:            5633.58$\pm$1.77 & 36 & 2.632$\pm$0.010   & QSO \\
                 & C{\sc iii}] 1909:          6922.36$\pm$4.56 & 49 & (2.68$\pm$0.11)   & \\
STR 1\_1 048     & Si{\sc iv}/O{\sc iv}\,1397:5074.88$\pm$3.16 & 11 & 2.629$\pm$0.009   & QSO \\
                 & C{\sc iii}] 1909:          6918.14$\pm$6.02 & 15 &                   & \\
STR 1\_1 106     & Si{\sc iv}/O{\sc iv}\,1397:5019.54$\pm$1.37 & 16 & 2.584$\pm$0.018   & QSO \\
                 & C{\sc iii}] 1909:          6824.13$\pm$2.93 & 23 &                   & \\
LMC 4\_2 038     & C{\sc ii} 1335:            4716.57$\pm$2.67 & 14 & 2.529$\pm$0.005   & QSO \\
                 & Si{\sc iv}/O{\sc iv}\,1397:4936.46$\pm$0.77 & 10 & (2.54$\pm$0.15)   & \\
                 & C{\sc iv} 1549:            5464.78$\pm$1.14 &  6 &                   & \\
                 & C{\sc iii}] 1909:          6723.00$\pm$1.58 & 23 &                   & \\
BRI 3\_7 019     & Si{\sc iv}/O{\sc iv}\,1397:4913.21$\pm$2.16 & 21 & 2.515$\pm$0.002   & QSO \\
                 & C{\sc iv} 1549:            5444.26$\pm$0.75 & 17 & (2.55$\pm$0.08)   & \\
                 & C{\sc iii}] 1909:          6708.01$\pm$2.37 & 20 &                   & \\
VMC Bright 24    & C{\sc iv} 1549:            5443.11$\pm$0.96 & 13 & 2.513$\pm$0.001   & QSO \\  
                 & C{\sc iii}] 1909:          6704.66$\pm$2.10 & 12 & (2.51$\pm$0.11)   & \\
STR 2\_1 112     & Si{\sc iv}/O{\sc iv}\,1397:4829.83$\pm$0.93 & 26 & 2.453$\pm$0.005   & QSO \\
                 & C{\sc iv} 1549:            5346.76$\pm$0.37 & 18 &                   & \\
                 & C{\sc iii}] 1909:          6582.84$\pm$1.52 & 15 &                   & \\
BRI 3\_7 182     & Si{\sc iv}/O{\sc iv}\,1397:4655.80$\pm$1.45 & 31 & 2.329$\pm$0.005   & QSO \\
                 & C{\sc iv} 1549:            5148.37$\pm$2.53 & 20 & (2.33$\pm$0.32)   & \\
                 & C{\sc iii}] 1909:          6354.32$\pm$4.80 &  5 &                   & \\
SMC 4\_2 015     & C{\sc iv} 1549:            5151.69$\pm$2.06 & 20 & 2.324$\pm$0.003   & QSO \\
                 & C{\sc iii}] 1909:          6342.21$\pm$6.72 & 11 & (2.43$\pm$0.05)   & \\
STR 1\_1 083     & C{\sc iv} 1549:            5052.71$\pm$1.19 & 18 & 2.264$\pm$0.002   & QSO \\
                 & He{\sc ii} 1640:           5352.79$\pm$1.04 & 14 &                   & \\
                 & C{\sc iii}] 1909:          6234.80$\pm$2.84 & 22 &                   & \\
STR 1\_1 034     & C{\sc iv} 1549:            5033.76$\pm$0.69 & 38 & 2.250$\pm$0.002   & QSO \\
                 & He{\sc ii} 1640:           5333.90$\pm$0.29 & 23 &                   & \\
                 & C{\sc iii}] 1909:          6199.80$\pm$2.39 & 43 &                   & \\
SMC 6\_3 161     & C{\sc iv} 1549:            4971.86$\pm$1.27 &  4 & 2.207$\pm$0.004   & QSO \\
                 & C{\sc iii}] 1909:          6117.80$\pm$1.57 &  7 &                   & \\
SMC 3\_1 035     & C{\sc iv} 1549:            4966.84$\pm$1.38 & 26 & 2.204$\pm$0.004   & QSO \\
                 & C{\sc iii}] 1909:          6112.70$\pm$1.53 & 74 &                   & \\
BRI 2\_7 021     & C{\sc iv} 1549:            4965.34$\pm$6.05 & 12 & 2.199$\pm$0.013   & QSO \\
                 & C{\sc iii}] 1909:          6094.28$\pm$4.82 & 16 &                   & \\
SMC 6\_5 077     & C{\sc iv} 1549:            4939.35$\pm$1.10 & 29 & 2.188$\pm$0.001   & QSO \\
                 & C{\sc iii}] 1909:          6083.85$\pm$1.94 & 10 & (2.19$\pm$0.06)   & \\
\hline
\end{tabular}
\end{small}
\end{center}
\end{table}
\addtocounter{table}{-1}

\begin{table}[]
\caption{Continued.}
\begin{center}
\begin{small}
\begin{tabular}{@{}l@{}c@{}c@{}c@{}c@{}}
\hline\hline
Object  ID       & Spectral feature, observed & ~S/N~  &~~~Redshift~~~~~~       & Class.\\
                 & central wavelength (\AA)   &        & $z$                    & \\
\hline
BRI 3\_4 003     & C{\sc iv} 1549:            4940.52$\pm$1.25 & 44 & 2.187$\pm$0.004   & QSO \\
                 & C{\sc iii}] 1909:          6079.35$\pm$1.85 & 31 & (1.77$\pm$0.52)   & \\
LMC 3\_5 091     & C{\sc iv} 1549:    4929.45$\pm$3.22  &  7 & 2.184$\pm$0.002   & QSO \\
                 & C{\sc iii}] 1909:  6079.79$\pm$2.97  &  8 &                   & \\
                 & [Ne{\sc iv}] 2424: 7723.13$\pm$0.11  &  3 &                   & \\
BRI 2\_3 007     & C{\sc iv} 1549:    4931.77$\pm$0.58  & 35 & 2.179$\pm$0.009   & QSO \\
                 & C{\sc iii}] 1909:  6059.45$\pm$6.27  & 27 & (2.18$\pm$0.06)   & \\
BRI 3\_4 141     & C{\sc iv} 1549:    4914.75$\pm$2.30  & 10 & 2.169$\pm$0.007   & QSO \\
                 & C{\sc iii}] 1909:  6042.78$\pm$3.33  & 21 & (2.10$\pm$0.16)   & \\
BRI 3\_6 031     & C{\sc iv} 1549:    4911.88$\pm$0.66  &  5 & 2.167$\pm$0.008   & QSO \\
                 & C{\sc iii}] 1909:  6037.71$\pm$4.59  & 12 &                   & \\
BRI 3\_7 127     & He{\sc ii} 1640:   5155.00$\pm$0.57  &  9 & 2.142$\pm$0.002   & QSO \\
                 & C{\sc iii}] 1909:  5994.67$\pm$1.04  & 14 &                   & \\
STR 1\_1 108     & C{\sc iv} 1549:    4835.49$\pm$1.57  &  5 & 2.113$\pm$0.010   & QSO \\
                 & C{\sc iii}] 1909:  5921.90$\pm$1.17  &  9 &                   & \\
                 & Mg{\sc ii} 2798:   8715.96$\pm$1.56  &  5 &                   & \\
LMC 4\_2 120     & C{\sc iv} 1549:    4789.34$\pm$2.16  & 49 & 2.093$\pm$0.008   & QSO \\
                 & C{\sc iii}] 1909:  5889.97$\pm$8.39  & 20 & (2.10$\pm$0.09)   & \\
                 & Mg{\sc ii} 2798:   8679.38$\pm$2.18  & 11 &                   & \\
BRI 2\_4 025     & C{\sc iv} 1549:    4767.27$\pm$1.09  & 72 & 2.089$\pm$0.010   & QSO \\
                 & C{\sc iii}] 1909:  5908.42$\pm$1.86  & 31 & (2.09$\pm$0.09)   & \\
                 & Mg{\sc ii} 2798:   8661.97$\pm$0.83  & 13 &                   & \\
BRI 2\_3 069     & C{\sc iv} 1549:    4781.22$\pm$1.32  & 72 & 2.089$\pm$0.005   & QSO \\
                 & C{\sc iii}] 1909:  5889.74$\pm$2.92  & 31 & (2.10$\pm$0.07)   & \\
                 & Mg{\sc ii} 2798:   8661.86$\pm$0.71  & 13 &                   & \\
LMC 4\_2 033g    & C{\sc iv} 1549:    4742.68$\pm$0.37  & 65 & 2.0620$\pm$0.0005 & QSO \\
                 & He{\sc ii} 1640:   5022.67$\pm$3.84  & 53 & (2.08$\pm$0.12)   & \\
                 & C{\sc iii}] 1909:  5845.69$\pm$0.66  &  8 &                   & \\
SMC 3\_1 119     & C{\sc iii}] 1909:  5811.64$\pm$3.13  & 20 & 2.047$\pm$0.005   & QSO \\
                 & Mg{\sc ii} 2798:   8536.43$\pm$4.52  & 22 & (1.17$\pm$0.60)   & \\
STR 2\_1 055     & C{\sc iv} 1549:    4706.16$\pm$0.34  & 46 & 2.041$\pm$0.004   & QSO \\
                 & C{\sc iii}] 1909:  5800.23$\pm$1.11  & 27 & (2.04$\pm$0.12)   & \\
                 & Mg{\sc ii} 2798:   8521.77$\pm$1.66  & 23 &                   & \\
STR 2\_1 100     & C{\sc iv} 1549:    4701.98$\pm$0.47  & 36 & 2.0354$\pm$0.0006 & QSO \\
                 & C{\sc iii}] 1909:  5792.56$\pm$0.67  & 20 & (2.83$\pm$0.11)   & \\
                 & Mg{\sc ii} 2798:   8497.14$\pm$0.85  & 26 &                   & \\
SMC 5\_3 089     & C{\sc iv} 1549:    4680.37$\pm$0.60  & 10 & 2.028$\pm$0.008   & QSO \\
                 & C{\sc iii}] 1909:  5778.01$\pm$1.91  & 18 & (2.02$\pm$0.17)   & \\
                 & Mg{\sc ii} 2798:   8498.26$\pm$0.86  & 12 &                   & \\
BRI 2\_7 151     & C{\sc iii}] 1909:  5758.76$\pm$4.70  & 31 & 2.024$\pm$0.014   & QSO \\
                 & Mg{\sc ii} 2798:   8483.91$\pm$1.27  & 12 & (1.66$\pm$0.53)   & \\
BRI 3\_6 100     & C{\sc iv} 1549:    4683.10$\pm$5.68  & 10 & 2.024$\pm$0.002   & QSO \\
                 & C{\sc iii}] 1909:  5773.53$\pm$3.05  & 15 &                   & \\
SMC 5\_3 075     & C{\sc iv} 1549:    4667.79$\pm$0.65  &  7 & 2.014$\pm$0.001   & QSO \\
                 & C{\sc iii}] 1909:  5752.02$\pm$2.03  &  3 &                   & \\
                 & Mg{\sc ii} 2798:   8439.86$\pm$0.31  &  3 &                   & \\
SMC 6\_5 080     & C{\sc iv} 1549:    4629.27$\pm$2.49  & 11 & 1.993$\pm$0.008   & QSO \\
                 & C{\sc iii}] 1909:  5704.28$\pm$3.14  & 30 & (2.00$\pm$0.07)   & \\
                 & Mg{\sc ii} 2798:   8401.87$\pm$1.13  & 12 &                   & \\
SMC 5\_3 128     & C{\sc iv} 1549:    4588.67$\pm$0.50  & 20 & 1.962$\pm$0.006   & QSO \\
                 & C{\sc iii}] 1909:  5641.74$\pm$3.09  & 13 & (1.96$\pm$0.13)   & \\
                 & Mg{\sc ii} 2798:   8305.60$\pm$4.97  & 24 &                   & \\
BRI 2\_7 130     & C{\sc iii}] 1909:  5621.91$\pm$0.64  & 25 & 1.948$\pm$0.006   & QSO \\
                 & Mg{\sc ii} 2798:   8259.85$\pm$10.88 & 10 &                   & \\
BRI 2\_7 145     & C{\sc iii}] 1909:  5616.86$\pm$1.35  & 23 & 1.941$\pm$0.002   & QSO \\
                 & Mg{\sc ii} 2798:   8228.99$\pm$5.94  & 21 & (2.71$\pm$0.37)   & \\
SMC 5\_3 015     & C{\sc iii}] 1909:  5595.05$\pm$1.63  &  8 & 1.936$\pm$0.009   & QSO \\
                 & Mg{\sc ii} 2798:   8228.94$\pm$5.92  & 13 & (1.95$\pm$0.08)   & \\
BRI 2\_4 121     & C{\sc iii}] 1909:  5479.44$\pm$0.96  & 26 & 1.876$\pm$0.011   & QSO \\
                 & Mg{\sc ii} 2798:   8064.67$\pm$1.23  & 14 & (1.88$\pm$0.08)   & \\
SMC 4\_5 060     & C{\sc iii}] 1909:  5480.88$\pm$0.36  & 14 & 1.876$\pm$0.009   & QSO \\
                 & Mg{\sc ii} 2798:   8061.91$\pm$0.89  & 18 & (1.91$\pm$0.54)   & \\
BRI 3\_4 004     & C{\sc iii}] 1909:  5403.63$\pm$2.40  & 23 & 1.830$\pm$0.003   & QSO \\
                 & Mg{\sc ii} 2798:   7915.43$\pm$6.09  &  3 &                   & \\
\hline
\end{tabular}
\end{small}
\end{center}
\end{table}
\addtocounter{table}{-1}

\begin{table}[]
\caption{Continued.}
\begin{center}
\begin{small}
\begin{tabular}{@{}l@{}c@{}c@{}c@{}c@{}}
\hline\hline
Object  ID       & Spectral feature, observed & ~S/N~  &~~~Redshift~~~~~~       & Class.\\
                 & central wavelength (\AA)   &        & $z$                    & \\
\hline
LMC 3\_5 078     & C{\sc iii}] 1909:  5392.65$\pm$2.02  & 54 & 1.828$\pm$0.005   & QSO \\
                 & Mg{\sc ii} 2798:   7921.67$\pm$0.79  & 27 & (1.83$\pm$0.09)   & \\
SMC 4\_5 038     & C{\sc iii}] 1909:  5354.78$\pm$3.87  & 28 & 1.813$\pm$0.015   & QSO \\
                 & Mg{\sc ii} 2798:   7893.77$\pm$2.06  & 24 & (1.82$\pm$0.04)   & \\
SMC 6\_5 052     & C{\sc iii}] 1909:  5343.28$\pm$2.19  & 15 & 1.799$\pm$0.001   & QSO \\
                 & Mg{\sc ii} 2798:   7832.78$\pm$3.11  &  5 & (2.60$\pm$0.34)   & \\
SMC 3\_1 107     & C{\sc iii}] 1909:  5250.34$\pm$0.86  & 20 & 1.751$\pm$0.015   & QSO \\
                 &                                      &    & (1.78$\pm$0.07)   & \\
BRI 3\_4 129     & C{\sc iii}] 1909:  5236.27$\pm$0.94  & 10 & 1.743$\pm$0.015   & QSO \\
                 &                                      &    & (1.69$\pm$0.12)   & \\
SMC 3\_1 043     & C{\sc iii}] 1909:  5207.03$\pm$0.69  & 18 & 1.728$\pm$0.015   & QSO \\
                 &                                      &    & (1.72$\pm$0.09)   & \\
STR 1\_1 047     & C{\sc iii}] 1909:  5153.30$\pm$1.96  & 48 & 1.702$\pm$0.005   & QSO \\
                 & C{\sc ii}] 2326:   6292.63$\pm$2.55  &  7 & (1.70$\pm$0.11)   & \\
SMC 3\_1 020     & C{\sc iii}] 1909:  5154.03$\pm$0.49  & 22 & 1.700$\pm$0.015   & QSO \\
                 &                                      &    & (1.70$\pm$0.06)   & \\
BRI 3\_4 125     & C{\sc iii}] 1909:  5140.55$\pm$2.40  & 14 & 1.693$\pm$0.015   & QSO \\
                 &                                      &    & (1.35$\pm$0.25)   & \\
LMC 4\_2 066     & C{\sc iii}] 1909:  5058.30$\pm$1.67  & 34 & 1.652$\pm$0.004   & QSO \\
                 & C{\sc ii}] 2326:   6165.69$\pm$6.36  & 49 & (1.65$\pm$0.09)   & \\
                 & Mg{\sc ii} 2798:   7436.28$\pm$1.04  & 26 &                   & \\
BRI 2\_3 159     & Mg{\sc ii} 2798:   7149.17$\pm$13.13 & 15 & 1.552$\pm$0.006   & QSO \\
                 & C{\sc iii}] 1909:  4864.56$\pm$0.79  & 40 & (1.57$\pm$0.07)   & \\
STR 2\_1 169     & C{\sc iii}] 1909:  4865.72$\pm$1.16  & 32 & 1.550$\pm$0.003   & QSO \\
                 & Mg{\sc ii} 2798:   7141.97$\pm$0.94  & 12 & (1.56$\pm$0.09)   & \\
STR 1\_1 046     & C{\sc iii}] 1909:  4728.13$\pm$2.53  & 41 & 1.480$\pm$0.006   & QSO \\
                 & C{\sc ii}] 2326:   5761.64$\pm$4.97  & 24 & (1.49$\pm$0.08)   & \\
                 & Mg{\sc ii} 2798:   6961.19$\pm$2.50  & 59 &                   & \\
SMC 6\_3 229     & C{\sc iii}] 1909:  4653.95$\pm$6.38  &  9 & 1.439$\pm$0.002   & QSO \\
                 & Mg{\sc ii} 2798:   6829.42$\pm$0.43  & 16 & (1.40$\pm$0.29)   & \\
BRI 2\_7 018     & Mg{\sc ii} 2798:   6651.72$\pm$2.48  & 12 & 1.377$\pm$0.015   & QSO \\
                 &                                      &    & (1.38$\pm$0.07)   & \\
STR 1\_1 072g    & Mg{\sc ii} 2798:   6645.87$\pm$0.56  & 16 & 1.375$\pm$0.015:  & QSO \\
BRI 2\_3 011     & Mg{\sc ii} 2798:   6640.23$\pm$1.81  &  5 & 1.3728$\pm$0.0005 & QSO \\
                 & [Ne{\sc v}] 3426:  8132.13$\pm$0.07  & 25 & (1.98$\pm$0.10)   & \\
BRI 2\_4 150     & Mg{\sc ii} 2798:   6574.30$\pm$0.16  & 12 & 1.349$\pm$0.015   & QSO \\
BRI 2\_7 042g    & Mg{\sc ii} 2798:   6543.83$\pm$1.20  &  7 & 1.338$\pm$0.001:  & QSO \\
                 & [O{\sc ii}] 3727:  8714.09$\pm$0.87  & 14 & (1.33$\pm$0.32)   & \\
LMC 8\_3 039     & Mg{\sc ii} 2798:   6529.01$\pm$0.87  & 20 & 1.333$\pm$0.015   & QSO \\
                 &                                      &    & (1.33$\pm$0.07)   & \\
BRI 2\_3 143     & Mg{\sc ii} 2798:   6424.49$\pm$1.57  & 16 & 1.295$\pm$0.015   & QSO \\
                 &                                      &    & (1.67$\pm$0.10)   & \\
STR 1\_1 128     & Mg{\sc ii} 2798:   6381.43$\pm$0.32  & 39 & 1.280$\pm$0.015   & QSO \\
                 &                                      &    & (1.73$\pm$0.52)   & \\
LMC 3\_5 103     & Mg{\sc ii} 2798:   6317.76$\pm$3.50  &  3 & 1.257$\pm$0.015:  & QSO \\
STR 2\_1 117     & Mg{\sc ii} 2798:   6225.37$\pm$1.02  & 15 & 1.2243$\pm$0.0001 & QSO \\
                 & H$\epsilon$:       8832.79$\pm$0.45  &  6 & (1.26$\pm$0.08)   & \\
STR 2\_1 217g    & C{\sc ii}] 2326:   5163.17$\pm$0.18  &  8 & 1.220$\pm$0.001   & QSO \\
                 & [Ne{\sc iv}] 2424: 5377.69$\pm$0.12  & 10 & (0.96$\pm$0.33)   & \\
                 & Mg{\sc ii} 2798:   6217.89$\pm$1.69  &  9 &                   & \\
                 & [O{\sc ii}] 3727:  8272.05$\pm$0.10  & 13 &                   & \\
SMC 6\_3 141     & Mg{\sc ii} 2798:   6211.10$\pm$0.12  & 34 & 1.218$\pm$0.003   & QSO \\
                 & [Ne{\sc iv}] 2424: 5371.05$\pm$0.15  & 22 & (1.70$\pm$0.59)   & \\
BRI 3\_4 041g    & Mg{\sc ii} 2798:   6149.42$\pm$1.50  & 10 & 1.197$\pm$0.015:  & QSO \\
BRI 3\_6 145     & C{\sc ii}] 2326:   5062.73$\pm$1.63  & 43 & 1.175$\pm$0.001   & QSO \\
                 & Mg{\sc ii} 2798:   6085.37$\pm$0.70  & 29 & (1.69$\pm$0.39)   & \\
                 & [Ne{\sc v}] 3347:  7282.53$\pm$0.01  &  6 &                   & \\
BRI 2\_7 064     & Mg{\sc ii} 2798:   5966.32$\pm$2.00  & 24 & 1.132$\pm$0.015   & QSO \\
                 &                                      &    & (1.20$\pm$0.14)   & \\
BRI 2\_7 060     & Mg{\sc ii} 2798:   5960.43$\pm$0.45  & 14 & 1.130$\pm$0.015   & QSO \\
BRI 3\_4 157     & Mg{\sc ii} 2798:   5911.67$\pm$0.43  & 39 & 1.1120$\pm$0.0004 & QSO \\
                 & O{\sc iii} 3128:   6605.18$\pm$0.11  & 13 & (1.11$\pm$0.11)   & \\
\hline
\end{tabular}
\end{small}
\end{center}
\end{table}
\addtocounter{table}{-1}

\begin{table}[]
\caption{Continued.}
\begin{center}
\begin{small}
\begin{tabular}{@{}l@{}c@{}c@{}c@{}c@{}}
\hline\hline
Object  ID       & Spectral feature, observed & ~S/N~  &~~~Redshift~~~~~~       & Class.\\
                 & central wavelength (\AA)   &        & $z$                    & \\
\hline
BRI 3\_4 026     & [Ne{\sc iv}] 2424: 5119.62$\pm$2.06  & 15 & 1.111$\pm$0.001   & QSO \\
                 & Mg{\sc ii} 2798:   5911.25$\pm$0.13  & 45 & (1.50$\pm$0.13)   & \\
                 & H$\delta$:         8657.51$\pm$0.63  &  8 &                   & \\
BRI 3\_4 046     & Mg{\sc ii} 2798:   5894.68$\pm$0.22  &  4 & 1.106$\pm$0.015   & QSO \\
SMC 5\_6 057g    & Mg{\sc ii} 2798:   5852.98$\pm$4.33  & 24 & 1.091$\pm$0.015   & QSO \\
                 &                                      &    & (1.10$\pm$0.07)   & \\
BRI 3\_6 068     & [Ne{\sc iv}] 2424: 5065.43$\pm$0.32  & 34 & 1.090$\pm$0.006   & QSO \\
                 & Mg{\sc ii} 2798:   5847.26$\pm$1.03  & 21 & (1.09$\pm$0.17)   & \\
SMC 6\_3 199     & Mg{\sc ii} 2798:   5738.40$\pm$0.13  & 25 & 1.050$\pm$0.015   & QSO \\
BRI 2\_7 089     & Mg{\sc ii} 2798:   5707.13$\pm$0.18  & 10 & 1.039$\pm$0.015   & QSO \\
BRI 3\_4 166     & Mg{\sc ii} 2798:   5694.44$\pm$1.23  & 22 & 1.0353$\pm$0.0006 & QSO \\
                 & [Ne{\sc v}] 3426:  6975.10$\pm$0.67  &  7 &                   & \\
                 & H$\delta$:         8353.16$\pm$0.44  &  2 &                   & \\
BRI 3\_4 046\_2  & Mg{\sc ii} 2798:   5525.92$\pm$3.91  & 55 & 0.974$\pm$0.005   & QSO \\  
BRI 2\_3 057     & Mg{\sc ii} 2798:   4997.42$\pm$5.09  &  2 & 0.786$\pm$0.005   & QSO \\
VMC Bright 14    & O{\sc iii} 3128:    5578.65$\pm$0.02 &  3 & 0.776$\pm$0.009:  & QSO \\  
                 & [O{\sc ii}] 3727:   6628.50$\pm$0.81 & 12 &                   & \\
                 & H$\beta$:           8652.79$\pm$1.55 &  5 &                   & \\
                 & [O{\sc iii}] 5007:  8824.93$\pm$0.02 &  2 &                   & \\
SMC 6\_5 114     & Mg{\sc ii} 2798:    4857.81$\pm$0.13 & 19 & 0.735$\pm$0.001   & QSO \\
                 & H$\delta$:          7117.72$\pm$1.65 & 40 &                   & \\
                 & H$\gamma$:          7538.01$\pm$1.70 &  8 &                   & \\
                 & H$\beta$:           8441.16$\pm$0.72 & 28 &                   & \\
                 & [O{\sc iii}] 4959:  8599.12$\pm$0.43 & 27 &                   & \\
                 & [O{\sc iii}] 5007:  8685.61$\pm$0.31 & 13 &                   & \\
BRI 3\_4 109g    & Mg{\sc ii} 2798:    4847.47$\pm$0.41 & 50 & 0.732$\pm$0.001   & QSO \\
                 & [Ne{\sc v}] 3426:   5932.34$\pm$0.22 & 21 &                   & \\
                 & [O{\sc ii}] 3727:   6455.15$\pm$0.19 & 11 &                   & \\
                 & [Ne{\sc iii}] 3870: 6698.28$\pm$0.12 & 12 &                   & \\
                 & H$\delta$:          7101.05$\pm$0.47 & 29 &                   & \\
                 & H$\beta$:           8437.06$\pm$1.68 & 16 &                   & \\
                 & [O{\sc iii}] 4959:  8586.67$\pm$0.07 &  4 &                   & \\
                 & [O{\sc iii}] 5007:  8669.34$\pm$0.02 &  3 &                   & \\
SMC 6\_3 308     & H$\gamma$:          7300.67$\pm$1.27 & 12 & 0.6814$\pm$0.0004 & QSO \\
                 & H$\beta$:           8177.87$\pm$0.39 & 14 &                   & \\
                 & [O{\sc iii}] 5007:  8419.07$\pm$2.59 &  9 &                   & \\
BRI 3\_6 143     & Mg{\sc ii} 2798:    4670.93$\pm$0.78 & 28 & 0.669$\pm$0.001   & QSO \\
                 & O{\sc iii} 3128:    5220.71$\pm$0.68 & 57 &                   & \\
                 & H$\gamma$:          7250.01$\pm$1.57 & 18 &                   & \\
                 & H$\beta$:           8105.99$\pm$1.84 & 55 &                   & \\
                 & [O{\sc iii}] 4959:  8273.69$\pm$1.60 & 34 &                   & \\
                 & [O{\sc iii}] 5007:  8355.97$\pm$0.64 &  7 &                   & \\
BRI 3\_4 218g    & Mg{\sc ii} 2798:    4671.28$\pm$0.81 & 24 & 0.668$\pm$0.001   & QSO \\
                 & [Ne{\sc v}] 3426:   5708.37$\pm$0.50 & 27 &                   & \\
                 & [O{\sc ii}] 3727:   6221.21$\pm$0.09 & 17 &                   & \\
                 & [Ne{\sc iii}] 3870: 6451.71$\pm$1.10 & 17 &                   & \\
                 & H$\gamma$:          7250.83$\pm$0.85 & 29 &                   & \\
                 & H$\beta$:           8106.78$\pm$4.28 & 28 &                   & \\
                 & [O{\sc iii}] 4959:  8272.45$\pm$0.29 & 15 &                   & \\
                 & [O{\sc iii}] 5007:  8352.97$\pm$0.07 &  5 &                   & \\
BRI 3\_7 055g    & Mg{\sc ii} 2798:    4643.33$\pm$0.72 & 23 & 0.660$\pm$0.002   & QSO \\
                 & O{\sc iii} 3128:    5203.84$\pm$0.22 & 12 &                   & \\
                 & H$\beta$:           8062.31$\pm$1.11 & 22 &                   & \\
                 & [O{\sc iii}] 4959:  8233.97$\pm$0.45 & 17 &                   & \\
                 & [O{\sc iii}] 5007:  8311.05$\pm$0.15 & 13 &                   & \\
BRI 2\_3 238g    & [O{\sc ii}] 3727:   6140.92$\pm$0.03 &  5 & 0.6467$\pm$0.0003 & QSO \\
                 & H$\gamma$:          7151.32$\pm$1.54 & 14 &                   & \\
                 & H$\beta$:           8005.79$\pm$0.42 &  5 &                   & \\
                 & [O{\sc iii}] 4959:  8167.19$\pm$0.20 &  5 &                   & \\
                 & [O{\sc iii}] 5007:  8246.53$\pm$0.05 &  3 &                   & \\
BRI 3\_4 153     & [O{\sc ii}] 3727:   6140.60$\pm$0.54 & 11 & 0.646$\pm$0.001   & QSO \\
                 & H$\gamma$:          7141.50$\pm$5.15 & 15 &                   & \\
                 & H$\beta$:           7997.97$\pm$1.61 & 10 &                   & \\
\hline
\end{tabular}
\end{small}
\end{center}
\end{table}
\addtocounter{table}{-1}

\begin{table}[]
\caption{Continued.}
\begin{center}
\begin{small}
\begin{tabular}{@{}l@{}c@{}c@{}c@{}c@{}}
\hline\hline
Object  ID       & Spectral feature, observed & ~S/N~  &~~~Redshift~~~~~~       & Class.\\
                 & central wavelength (\AA)   &        & $z$                    & \\
\hline
SMC 3\_1 074g    & H$\beta$:           7992.96$\pm$0.39 & 16 & 0.644$\pm$0.005   & QSO \\
BRI 3\_4 190g    & H$\beta$:           7883.20$\pm$1.16 & 14 & 0.6221$\pm$0.0008:& QSO \\
                 & [O{\sc iii}] 4959:  8047.99$\pm$0.50 &  5 &                   & \\
                 & [O{\sc iii}] 5007:  8126.91$\pm$0.89 & 11 &                   & \\
BRI 3\_4 050g    & H$\gamma$:          7041.83$\pm$1.66 & 15 & 0.619$\pm$0.002   & QSO \\
                 & H$\beta$:           7865.06$\pm$2.34 &  3 &                   & \\
                 & [O{\sc iii}] 4959:  8026.24$\pm$0.60 &  3 &                   & \\
                 & [O{\sc iii}] 5007:  8102.59$\pm$0.45 &  6 &                   & \\
BRI 3\_4 033     & H$\delta$:          6608.25$\pm$0.78 & 13 & 0.6111$\pm$0.0004 & QSO \\
                 & H$\gamma$:          6997.52$\pm$0.69 & 15 &                   & \\
                 & H$\beta$:           7833.68$\pm$0.43 &  9 &                   & \\
                 & [O{\sc iii}] 4959:  7989.89$\pm$0.45 &  5 &                   & \\
                 & [O{\sc iii}] 5007:  8070.12$\pm$0.12 & 12 &                   & \\
BRI 3\_4 117     & [O{\sc ii}] 3727:   5989.23$\pm$0.38 & 22 & 0.607$\pm$0.002   & QSO \\
                 & [Ne{\sc iii}] 3870: 6213.84$\pm$0.16 & 12 & (0.60$\pm$0.11)   & \\
                 & H$\delta$:          6601.33$\pm$1.15 & 13 &                   & \\
                 & H$\gamma$:          6983.79$\pm$1.07 & 55 &                   & \\
                 & H$\beta$:           7812.20$\pm$1.35 &  8 &                   & \\
                 & [O{\sc iii}] 4959:  7959.57$\pm$1.07 & 11 &                   & \\
                 & [O{\sc iii}] 5007:  8043.32$\pm$0.84 &  8 &                   & \\
SMC 5\_6 047g    & H$\beta$:           7812.18$\pm$3.35 &  4 & 0.6067$\pm$0.0001 & QSO \\
                 & [O{\sc iii}] 4959:  7969.20$\pm$0.16 &  8 &                   & \\
                 & [O{\sc iii}] 5007:  7812.18$\pm$3.35 &  4 &                   & \\
BRI 2\_4 014g    & [O{\sc ii}] 3727:   5972.28$\pm$0.35 &  8 & 0.6013$\pm$0.0004:& QSO \\
                 & H$\beta$:           7785.60$\pm$2.99 &  1 &                   & \\
                 & [O{\sc iii}] 5007:  8018.89$\pm$0.42 &  5 &                   & \\
BRI 3\_7 043     & O{\sc iii} 3128:    4999.77$\pm$0.59 &  6 & 0.595$\pm$0.002   & QSO \\
                 & H$\delta$:          6539.82$\pm$0.67 &  3 &                   & \\
                 & H$\gamma$:          6932.02$\pm$1.30 & 10 &                   & \\
                 & H$\beta$:           7755.73$\pm$0.79 &  3 &                   & \\
                 & [O{\sc iii}] 4959:  7906.86$\pm$0.17 &  3 &                   & \\
                 & [O{\sc iii}] 5007:  7985.43$\pm$0.20 &  4 &                   & \\
BRI 3\_4 271g    & [Ne{\sc v}] 3426:   5448.98$\pm$0.24 &  9 & 0.5905$\pm$0.0003 & QSO \\
                 & [O{\sc ii}] 3727:   5931.60$\pm$2.74 &  6 &                   & \\
                 & [Ne{\sc iii}] 3870: 6155.60$\pm$0.32 &  7 &                   & \\
                 & [O{\sc iii}] 4959:  7888.48$\pm$0.70 &  5 &                   & \\
                 & [O{\sc iii}] 5007:  7965.72$\pm$0.02 &  3 &                   & \\
STR 2\_1 174g    & [Ne{\sc iii}] 3870: 6022.87$\pm$0.30 & 21 & 0.5559$\pm$0.0007 & QSO \\
                 & H$\gamma$:          6758.78$\pm$0.69 & 24 & (0.55$\pm$0.08)   & \\
                 & [O{\sc iii}] 4959:  7714.34$\pm$0.21 &  5 &                   & \\
                 & [O{\sc iii}] 5007:  7789.72$\pm$0.07 &  8 &                   & \\
VMC Bright 11    & Mg{\sc ii} 2798:    4329.17$\pm$1.85 &  7 & 0.547$\pm$0.001   & QSO \\  
                 & [O{\sc iii}] 5007:  7749.34$\pm$0.01 &  2 & (0.54$\pm$0.09)   & \\
LMC 3\_5 063g    & O{\sc iii} 3128:    4793.22$\pm$0.55 &  8 & 0.533$\pm$0.003:  & QSO \\
                 & [Ne{\sc v}] 3347:   5127.04$\pm$0.63 & 10 & (0.52$\pm$0.09)   & \\
                 & [Ne{\sc v}] 3426:   5249.99$\pm$0.28 & 31 &                   & \\
                 & H$\beta$:           7478.66$\pm$4.07 &  7 &                   & \\
                 & [O{\sc iii}] 5007:  7672.26$\pm$0.13 &  9 &                   & \\
STR 2\_1 077g    & [Ne{\sc v}] 3426:   5253.98$\pm$0.16 &  8 & 0.5323$\pm$0.0007 & QSO \\
                 & [O{\sc ii}] 3727:   5712.90$\pm$0.12 &  4 & (0.54$\pm$0.14)   & \\
                 & [Ne{\sc iii}] 3870: 5930.38$\pm$0.59 &  3 &                   & \\
                 & H$\beta$:           7445.54$\pm$0.49 & 11 &                   & \\
                 & [O{\sc iii}] 5007:  7674.09$\pm$0.24 &  3 &                   & \\
STR 1\_1 100g    & [O{\sc ii}] 3727:   5686.24$\pm$0.12 &  6 & 0.5247$\pm$0.0004 & QSO \\
                 & H$\gamma$:          6619.65$\pm$0.47 & 17 & (0.53$\pm$0.06)   & \\
                 & H$\beta$:           7412.45$\pm$0.17 & 12 &                   & \\
BRI 2\_3 195g    & [O{\sc ii}] 3727:   5678.81$\pm$0.28 & 19 & 0.5228$\pm$0.0007 & QSO \\
                 & H$\epsilon$:        6043.35$\pm$0.44 &  6 & (0.54$\pm$0.09)   & \\
                 & H$\delta$:          6250.77$\pm$0.81 & 12 &                   & \\
                 & H$\gamma$:          6615.73$\pm$0.35 &  9 &                   & \\
                 & H$\beta$:           7405.91$\pm$0.60 &  4 &                   & \\
                 & [O{\sc iii}] 4959:  7551.90$\pm$0.30 &  6 &                   & \\
                 & [O{\sc iii}] 5007:  7622.94$\pm$0.31 &  3 &                   & \\
\hline
\end{tabular}
\end{small}
\end{center}
\end{table}
\addtocounter{table}{-1}

\begin{table}[]
\caption{Continued.}
\begin{center}
\begin{small}
\begin{tabular}{@{}l@{}c@{}c@{}c@{}c@{}}
\hline\hline
Object  ID       & Spectral feature, observed & ~S/N~  &~~~Redshift~~~~~~       & Class.\\
                 & central wavelength (\AA)   &        & $z$                    & \\
\hline
SMC 4\_2 071g    & [O{\sc iii}] 4959:  7583.59$\pm$0.15 &  4 & 0.5144$\pm$0.0004 & QSO \\
                 & [O{\sc iii}] 5007:  7512.75$\pm$0.38 &  6 &                   & \\
STR 1\_1 226g    & [Ne{\sc v}] 3426:   5171.38$\pm$0.11 & 16 & 0.5096$\pm$0.0003 & QSO \\
                 & [O{\sc ii}] 3727:   5629.77$\pm$0.05 &  7 & (0.52$\pm$0.07)   & \\
                 & H$\gamma$:          6554.22$\pm$0.64 & 23 &                   & \\
                 & H$\beta$:           7341.63$\pm$0.49 &  7 &                   & \\
                 & [O{\sc iii}] 4959:  7487.13$\pm$0.14 &  6 &                   & \\
                 & [O{\sc iii}] 5007:  7560.64$\pm$0.22 &  5 &                   & \\
SMC 6\_3 092g    & H$\delta$:          6135.86$\pm$0.79 &  6 & 0.494$\pm$0.001   & QSO \\
                 & H$\gamma$:          6494.21$\pm$0.37 &  5 &                   & \\
                 & H$\beta$:           7268.68$\pm$0.45 &  6 &                   & \\
                 & [O{\sc iii}] 4959:  7404.32$\pm$0.20 &  8 &                   & \\
                 & [O{\sc iii}] 5007:  7478.45$\pm$0.08 & 12 &                   & \\
SMC 6\_5 008g    & H$\delta$:                      6091.30$\pm$0.86 & 11 & 0.483$\pm$0.001   & QSO \\
                 & H$\gamma$:                      6445.26$\pm$0.67 & 17 & (0.45$\pm$0.07)   & \\
                 & H$\beta$:                       7211.77$\pm$0.28 & 32 &                   & \\
                 & [O{\sc iii}] 5007:              7420.58$\pm$0.48 & 11 &                   & \\
LMC 3\_5 085g    & [Ne{\sc v}] 3347:               4931.42$\pm$0.11 &  5 & 0.4731$\pm$0.0003 & QSO \\
                 & [Ne{\sc v}] 3426:               5049.54$\pm$0.03 &  2 &                   & \\
                 & [O{\sc ii}] 3727:               5492.53$\pm$0.02 &  1 &                   & \\
                 & H$\epsilon$:                    5847.30$\pm$0.05 &  2 &                   & \\
                 & H$\delta$:                      6044.63$\pm$0.06 &  4 &                   & \\
                 & H$\gamma$:                      6395.88$\pm$0.07 &  3 &                   & \\
                 & He{\sc ii} 4687:                6905.29$\pm$0.08 &  9 &                   & \\
                 & H$\beta$:                       7162.82$\pm$0.07 &  3 &                   & \\
                 & [O{\sc iii}] 4959:              7307.25$\pm$0.04 &  2 &                   & \\
                 & [O{\sc iii}] 5007:              7377.69$\pm$0.01 &  2 &                   & \\
STR 1\_1 097g    & [Ne{\sc v}] 3426:               4932.31$\pm$0.22 & 16 & 0.4394$\pm$0.0004 & QSO \\
                 & [O{\sc ii}] 3727:               5367.84$\pm$0.04 &  6 &                   & \\
                 & H$\gamma$:                      6252.96$\pm$0.76 & 12 &                   & \\
                 & H$\beta$:                       6998.56$\pm$0.24 & 11 &                   & \\
                 & [O{\sc iii}] 4959:              7138.35$\pm$0.08 &  8 &                   & \\
                 & [O{\sc iii}] 5007:              7207.47$\pm$0.02 &  4 &                   & \\
VMC Bright 05    & Mg{\sc ii} 2798:                4020.45$\pm$0.69 & 27 & 0.4360$\pm$0.0006 & QSO \\  
                 & [O{\sc ii}] 3727:               5357.62$\pm$0.08 & 22 & (0.44$\pm$0.18)   & \\
                 & [Ne{\sc iii}] 3870:             5556.50$\pm$0.19 & 17 &                   & \\
                 & H$\gamma$:                      6232.38$\pm$0.97 & 11 &                   & \\
                 & [O{\sc iii}] 4959:              6978.52$\pm$1.59 & 34 &                   & \\
                 & [O{\sc iii}] 5007:              7123.40$\pm$0.07 & 12 &                   & \\
                 & He{\sc i} 5877:                 7192.49$\pm$0.07 & 14 &                   & \\
BRI 3\_4 437g    & [O{\sc ii}] 3727:               5324.83$\pm$0.07 &  6 & 0.4285$\pm$0.0006 & QSO \\
                 & H$\beta$:                       6948.45$\pm$4.27 &  5 &                   & \\
                 & [O{\sc iii}] 4959:              7084.18$\pm$0.13 & 10 &                   & \\
                 & [O{\sc iii}] 5007:              7152.26$\pm$0.07 &  3 &                   & \\
                 & [Cl{\sc iii}]:/Fe{\sc ii}\,5539:7917.62$\pm$0.14 &  4 &                   & \\
SMC 6\_5 012g    & [Ne{\sc v}] 3347:               4780.72$\pm$0.29 &  4 & 0.4279$\pm$0.0008 & QSO \\
                 & [Ne{\sc v}] 3426:               4892.59$\pm$0.24 &  8 &                   & \\
                 & [O{\sc ii}] 3727:               5323.87$\pm$0.13 &  4 &                   & \\
                 & [Ne{\sc iii}] 3870:             5519.92$\pm$0.94 & 17 &                   & \\
                 & H$\beta$:                       6948.12$\pm$2.33 & 14 &                   & \\
                 & [O{\sc iii}] 4959:              7082.68$\pm$0.07 &  8 &                   & \\
                 & [O{\sc iii}] 5007:              7151.21$\pm$0.03 &  4 &                   & \\
STR 2\_1 240g    & [Ne{\sc v}] 3426:               4834.06$\pm$0.15 & 10 & 0.4102$\pm$0.0004 & QSO \\
                 & [O{\sc ii}] 3727:               5260.08$\pm$0.06 &  9 & (0.89$\pm$0.12)   & \\
                 & [Ne{\sc iii}] 3870:             5457.07$\pm$0.14 & 10 &                   & \\
                 & H$\delta$:                      5785.27$\pm$2.27 &  6 &                   & \\
                 & He{\sc ii} 4687:                6608.08$\pm$0.30 & 12 &                   & \\
                 & H$\beta$:                       6853.83$\pm$1.37 &  7 &                   & \\
                 & [O{\sc iii}] 4959:              6996.00$\pm$0.04 &  2 &                   & \\
                 & [O{\sc iii}] 5007:              7063.72$\pm$0.02 &  1 &                   & \\
                 & He{\sc i} 5877:                 8288.67$\pm$0.06 & 16 &                   & \\
\hline
\end{tabular}
\end{small}
\end{center}
\end{table}
\addtocounter{table}{-1}

\begin{table}[]
\caption{Continued.}
\begin{center}
\begin{small}
\begin{tabular}{@{}l@{}c@{}c@{}c@{}c@{}}
\hline\hline
Object  ID       & Spectral feature, observed & ~S/N~  &~~~Redshift~~~~~~       & Class.\\
                 & central wavelength (\AA)   &        & $z$                    & \\
\hline
VMC Bright 08    & H$\beta$:                       6731.92$\pm$0.82 &  7 & 0.383$\pm$0.001   & QSO \\  
                 & [O{\sc iii}] 4959:              6856.74$\pm$0.73 &  9 & (0.39$\pm$0.06)   & \\
                 & [O{\sc iii}] 5007:              6925.00$\pm$0.14 & 12 &                   & \\
STR 2\_1 203     & [Ne{\sc v}] 3426:               4594.59$\pm$0.10 & 13 & 0.3419$\pm$0.0008 & QSO \\
                 & [O{\sc ii}] 3727:               5004.20$\pm$0.06 &  5 & (0.46$\pm$0.10)   & \\
                 & [Ne{\sc iii}] 3870:             5191.98$\pm$0.07 &  9 &                   & \\
                 & H$\gamma$:                      5833.83$\pm$0.64 & 26 &                   & \\
                 & H$\beta$:                       6526.88$\pm$1.29 & 27 &                   & \\
                 & [O{\sc iii}] 4959:              6654.50$\pm$0.07 &  5 &                   & \\
                 & [O{\sc iii}] 5007:              6718.67$\pm$0.02 &  2 &                   & \\
                 & H$\alpha$:                      8809.18$\pm$0.26 & 11 &                   & \\
SMC 3\_1 044g    & H$\beta$:                       6517.27$\pm$2.37 &  9 & 0.3412$\pm$0.0008 & QSO \\
                 & [O{\sc iii}] 4959:              6654.96$\pm$0.12 &  6 & (0.34$\pm$0.06)   & \\
                 & [O{\sc iii}] 5007:              6719.03$\pm$0.04 &  4 &                   & \\
SMC 6\_5 086g    & [O{\sc iii}] 4959:              6652.09$\pm$0.17 &  7 & 0.3411$\pm$0.0001 & QSO \\
                 & [O{\sc iii}] 5007:              6716.39$\pm$0.21 &  9 &                   & \\
BRI 3\_7 176g    & H$\gamma$:          5761.37$\pm$1.78 &  6 & 0.3260$\pm$0.0007 & QSO \\
                 & H$\beta$:           6446.09$\pm$1.81 & 15 & (0.36$\pm$0.07)   & \\
                 & [O{\sc iii}] 4959:  6572.80$\pm$2.65 &  5 &                   & \\
                 & [O{\sc iii}] 5007:  6641.36$\pm$1.50 & 10 &                   & \\
                 & H$\alpha$:          8704.67$\pm$0.71 & 36 &                   & \\
BRI 3\_4 347g    & [O{\sc ii}] 3727:   4862.72$\pm$0.24 & 14 & 0.3039$\pm$0.0007 & QSO \\
                 & [Ne{\sc iii}] 3870: 5050.90$\pm$1.15 & 11 & (0.34$\pm$0.09)   & \\
                 & H$\delta$:          5348.44$\pm$0.53 & 24 &                   & \\
                 & H$\gamma$:          5663.61$\pm$1.07 & 17 &                   & \\
                 & H$\beta$:           6335.91$\pm$2.02 & 40 &                   & \\
                 & [O{\sc iii}] 4959:  6465.95$\pm$0.12 & 14 &                   & \\
                 & [O{\sc iii}] 5007:  6529.10$\pm$0.04 &  4 &                   & \\
                 & H$\alpha$:          8558.91$\pm$1.32 & 23 &                   & \\
BRI 3\_7 154g    & H$\epsilon$:        5165.96$\pm$0.44 & 13 & 0.3011$\pm$0.0004 & QSO \\
                 & H$\delta$:          5339.68$\pm$0.60 & 55 & (0.36$\pm$0.07)   & \\
                 & H$\gamma$:          5652.05$\pm$0.33 & 16 &                   & \\
                 & H$\beta$:           6326.07$\pm$0.27 & 14 &                   & \\
                 & [O{\sc iii}] 4959:  6451.30$\pm$0.40 & 23 &                   & \\
                 & [O{\sc iii}] 5007:  6516.31$\pm$0.13 & 14 &                   & \\
                 & H$\alpha$:          8539.48$\pm$0.12 & 14 &                   & \\
STR 2\_1 225g    & [O{\sc ii}] 3727:   4808.10$\pm$0.48 & 10 & 0.2889$\pm$0.0007 & QSO \\
                 & [O{\sc iii}] 5007:  6451.91$\pm$0.06 &  6 &                   & \\
                 & H$\alpha$:          8460.56$\pm$0.17 & 15 &                   & \\
SMC 6\_3 141\_2  & [O{\sc ii}] 3727:   4578.48$\pm$0.18 & 11 & 0.2278$\pm$0.0008 & QSO \\
                 & [O{\sc iii}] 5007:  6146.81$\pm$0.31 &  7 &                   & \\
                 & He{\sc i} 5877:     7224.64$\pm$7.83 & 11 &                   & \\
                 & H$\alpha$:          8056.47$\pm$0.06 &  2 &                   & \\
                 & [N{\sc ii}] 6586:   8082.92$\pm$0.17 & 15 &                   & \\
BRI 2\_7 053g    & H$\delta$:          4753.11$\pm$0.19 &  7 & 0.1584$\pm$0.0001 & QSO \\
                 & H$\gamma$:          5029.64$\pm$0.07 &  5 &                   & \\
                 & H$\beta$:           5632.63$\pm$0.03 &  2 &                   & \\
                 & [O{\sc iii}] 4959:  5745.84$\pm$0.01 &  4 &                   & \\
                 & [O{\sc iii}] 5007:  5801.45$\pm$0.04 &  1 &                   & \\
                 & He{\sc i} 5877:     6807.67$\pm$0.19 & 10 &                   & \\
                 & [O{\sc i}] 6303:    7299.90$\pm$0.13 & 14 &                   & \\
                 & [O{\sc i}] 6366:    7376.15$\pm$2.01 &  8 &                   & \\
                 & [S{\sc ii}] 6718:   7782.27$\pm$0.10 &  4 &                   & \\
                 & [S{\sc ii}] 6733:   7799.20$\pm$3.00 &  4 &                   & \\
BRI 3\_4 209g    & [O{\sc iii}] 4959:  5696.88$\pm$0.08 &  6 & 0.1485$\pm$0.0001 & QSO \\
                 & [O{\sc iii}] 5007:  5752.30$\pm$0.02 &  2 &                   & \\
                 & H$\alpha$:          7539.49$\pm$0.03 &  1 &                   & \\
                 & [S{\sc ii}] 6718:   7716.22$\pm$0.50 &  4 &                   & \\
                 & [S{\sc ii}] 6733:   7731.87$\pm$0.85 &  4 &                   & \\
                 & [N{\sc ii}] 6586:   7563.63$\pm$0.14 &  2 &                   & \\
                 & [O{\sc ii}] 7321:   8407.10$\pm$0.10 &  3 &                   & \\
\hline
\end{tabular}
\end{small}
\end{center}
\end{table}
\addtocounter{table}{-1}

\begin{table}[]
\caption{Continued.}
\begin{center}
\begin{small}
\begin{tabular}{@{}l@{}c@{}c@{}c@{}c@{}}
\hline\hline
Object  ID       & Spectral feature, observed & ~S/N~  &~~~Redshift~~~        & Class.\\
                 & central wavelength (\AA)   &        & $z$                  & \\
\hline
SMC 4\_2 014g    & H$\beta$:           5444.80$\pm$0.09 & 13 & 0.1195$\pm$0.0002 & QSO \\
                 & [O{\sc iii}] 4959:  5552.24$\pm$0.23 &  3 &                   & \\
                 & [O{\sc iii}] 5007:  5606.67$\pm$0.06 &  2 &                   & \\
                 & H$\alpha$:          7348.93$\pm$0.09 &  3 &                   & \\
BRI 3\_4 089     & H$\alpha$:          7332.89$\pm$0.48 &  1 & 0.117$\pm$0.005:  & QSO \\
STR 1\_1 033g    & H$\beta$:           5390.14$\pm$0.14 &  8 & 0.1084$\pm$0.0001 & QSO \\
                 & [O{\sc iii}] 4959:  5497.90$\pm$0.53 & 38 &                   & \\
                 & He{\sc i} 5877:     6513.99$\pm$0.55 & 29 &                   & \\
                 & H$\alpha$:          7276.64$\pm$0.48 &  2 &                   & \\
LMC 4\_2 043     & H$\alpha$:          7216.37$\pm$3.33 &  6 & 0.0993$\pm$0.005: & QSO \\
BRI 2\_3 032     & too low S/N                          &    &                   & unknown \\
BRI 2\_4 014     & too low S/N                          &    &                   & star?  \\
LMC 8\_3 013     & none                                 &    &                   & blue star \\
LMC 8\_3 016     & none                                 &    &                   & blue star \\
LMC 8\_3 023     & none                                 &    &                   & star \\
LMC 8\_3 095g    & none                                 &    &                   & blue star \\
SMC 4\_2 103     & too low S/N                          &    &                   & unknown \\
SMC 4\_5 062     & too low S/N                          &    &                   & unknown \\
SMC 5\_3 003g    & too low S/N                          &    &                   & unknown \\
SMC 5\_3 021     & Balmer lines, CaT                  &    &                   & star   \\
SMC 6\_3 186g    & too low S/N                        &    &                   & unknown \\
SMC 6\_3 195     & too low S/N                        &    &                   & unknown \\
SMC 6\_3 322     & too low S/N                        &    &                   & unknown \\
VMC Bright 02    & ~~~~no em. lines, poss. ell. gal.~~~&    & $\sim$0           & unknown \\  
VMC Bright 03    & too low S/N                        &    &                   & unknown \\  
\hline
\end{tabular}
\end{small}
\end{center}
\end{table}

\begin{figure*}
\centering
\includegraphics[width=18.4cm]{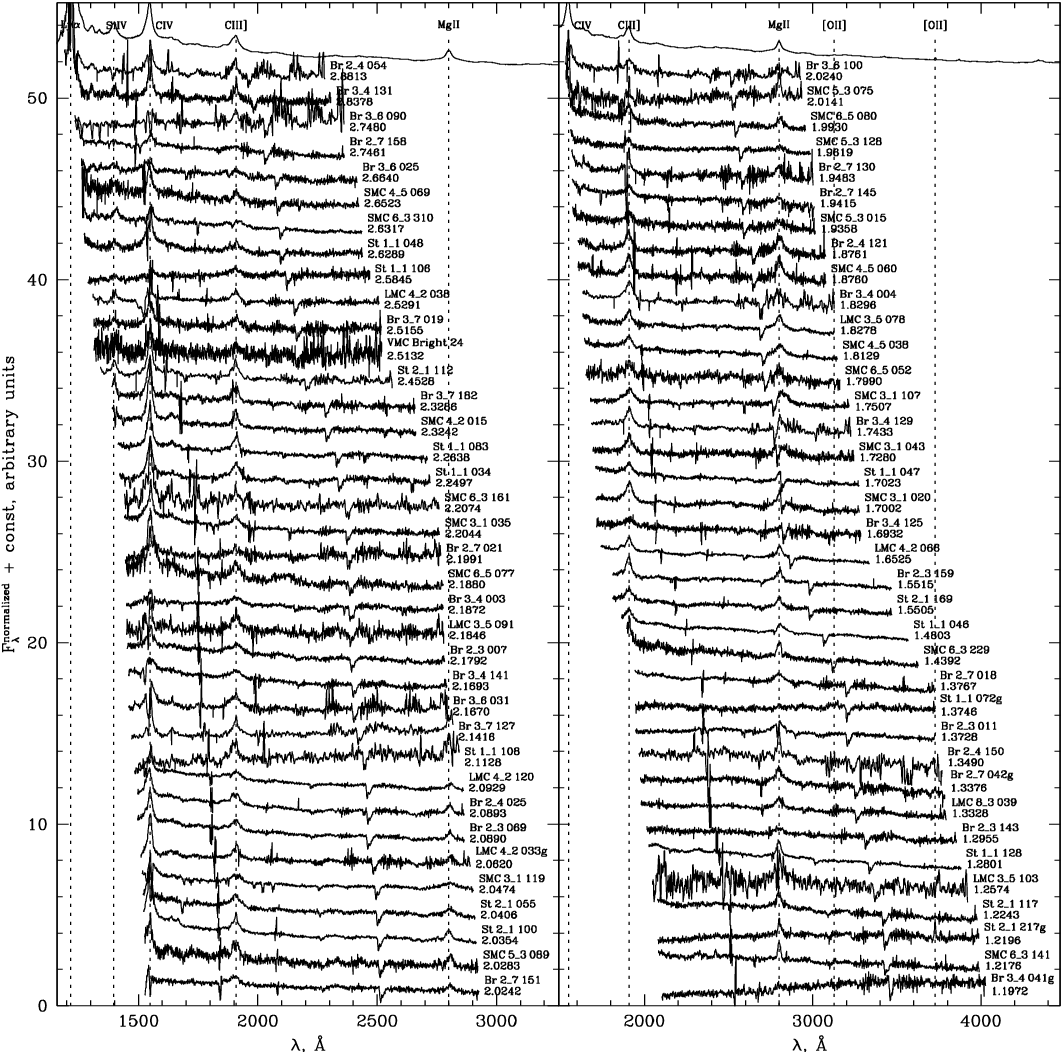}\\
\caption{Spectra of 148 objects (for three no spectra could be 
extracted, usually due to poor weather conditions) sorted by 
redshift and shifted to rest--frame wavelength. The spectra were 
normalized to an average value of one and shifted vertically by 
offsets of two, four, etc., for display purposes. The SDSS composite 
QSO spectrum \citep{2001AJ....122..549V} is shown at the top of all 
panels. 
Objects with no measured redshift due to lack of lines or low-S/N
are plotted assuming $z$=0 in the fifth panel next to the sky 
spectrum to facilitate the identification of the residuals from 
the sky emission lines.}\label{fig:spectra}
\end{figure*} 

\addtocounter{figure}{-1}
\begin{figure*}[!h]
\centering
\includegraphics[width=18.4cm]{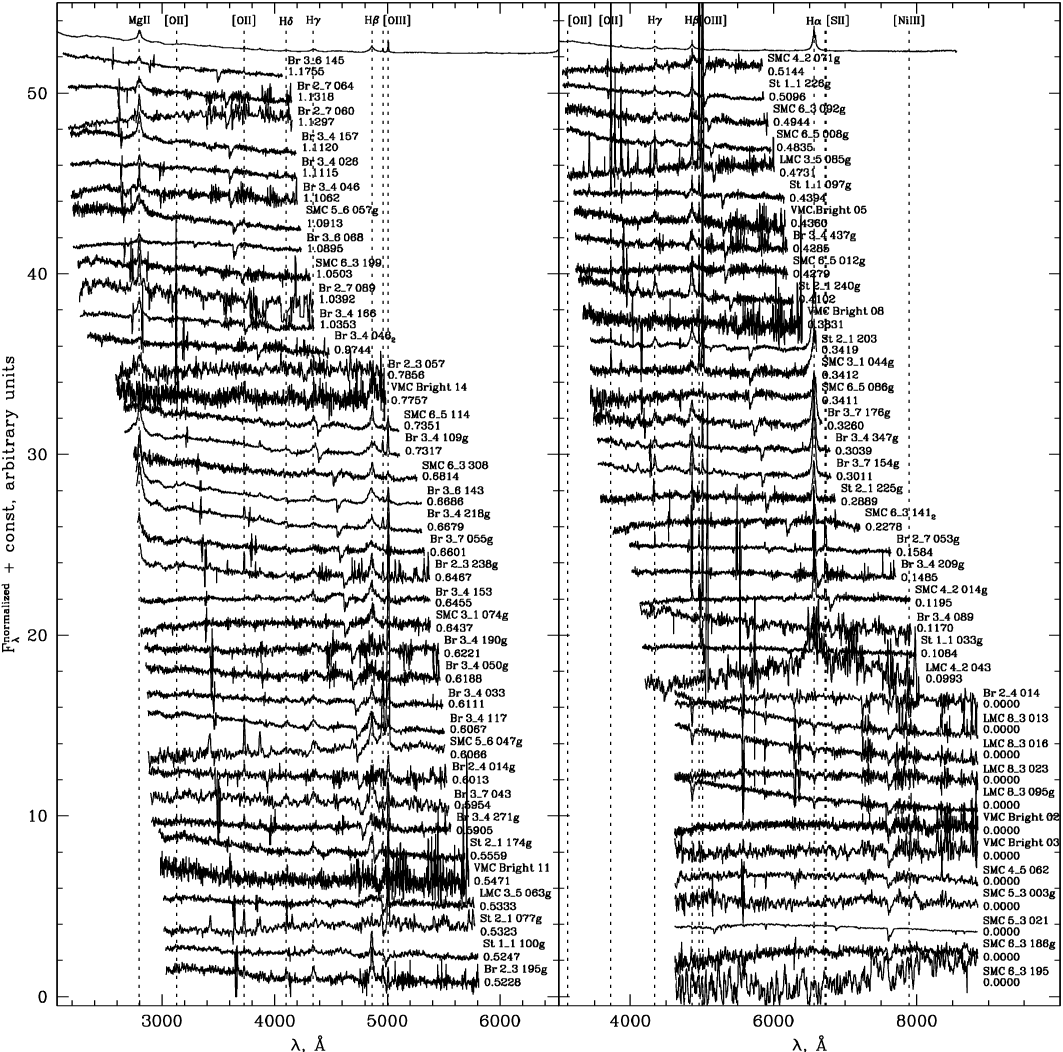} \\
\caption{Continued.}
\end{figure*}

\section{Results and Discussion}\label{sec:results}

\subsection{Quasar confirmation}

The majority of the candidates -- $136$ out of $148$ observed objects
-- are bona fide 
QSOs at $z$$\sim$0.1--2.9 (Fig.\,\ref{fig:spectra}). They all show 
some broad emission lines even though some spectra need block 
averaging, typically by 4 resolution bins, to make their features
clearly evident on the plot (the line measurements were done on the
original, not smoothed spectra). Ly$\alpha$ is visible in the spectra
of the highest redshift
QSOs, the rest of the sample shows other typical emission lines. 
The newly confirmed QSOs from the VLT observed sample are 
distributed as follows: $11$ in LMC, $34$ in SMC, $62$ in the Bridge 
and $24$ in the Stream areas; the five QSOs from the bright 
candidate list are: $1$ in LMC, $3$ in SMC and $1$ in the Bridge.
The spectra of six objects appear star-like with no broad emissions, 
and nine more are too noisy for secure classification.

The VDFS pipeline classifies 55 of our 151 objects as extended 
sources and 49 of these are spectroscopically confirmed QSOs. 
However, their extended nature seems to be more a matter of 
foreground contamination from Magellanic Cloud sources, rather than 
real resolving of the host galaxy, because 10 extended QSOs have 
redshift $z$$>$1 and their hosts are unlikely to be resolved under 
atmospheric seeing conditions. Furthermore, one of the extended 
non-QSOs is a blue LMC star and its extended nature is likely due 
to a surrounding star cluster. Therefore, the VDFS classification 
cannot be a critical constraint when a QSO candidate sample is 
assembled.

\begin{figure}
\centering
\includegraphics[width=8.75cm]{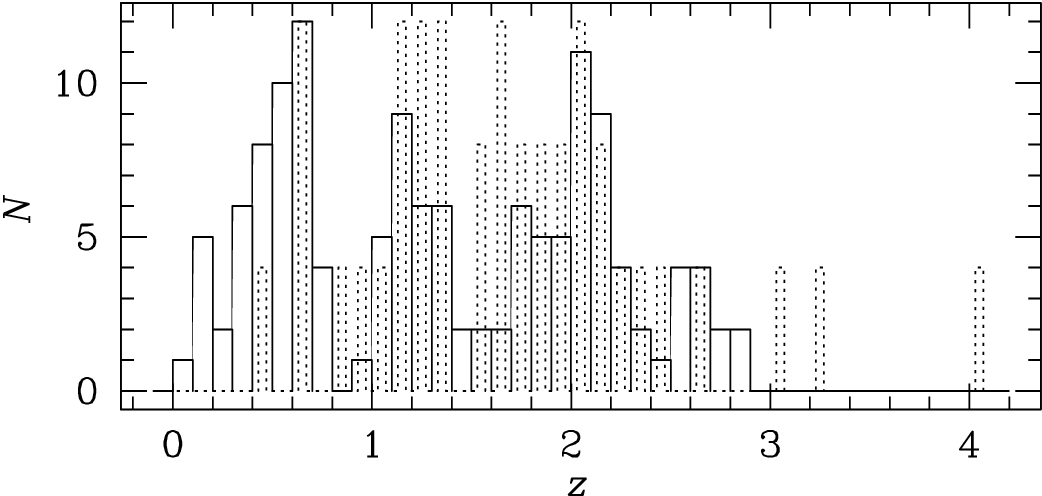}
\caption{Redshift histogram for 136 confirmed QSOs (solid line)
and for comparison the 47 objects with measured redshift from 
\citet[][dashed line]{2016A&A...588A..93I}.}\label{fig:z_hist}
\end{figure}

Figure\,\ref{fig:z_hist} shows that here we have a higher fraction
of confirmed QSOs with $z$$\leq$1 than \citet{2016A&A...588A..93I}.
This is likely due to the selection of brighter candidates and
because this work is based on a large number of Bridge tiles where
the background objects suffer less reddening and foreground
contamination than behind the LMC and SMC. Indeed, the colors of
lower redshift QSOs are closer to the stellar locus than those of
the higher redshift QSOs which makes them harder to identify in the
inner regions of the Magellanic Clouds and easier in the Bridge.

\begin{figure}[h!]
\centering
\includegraphics[width=9.0cm]{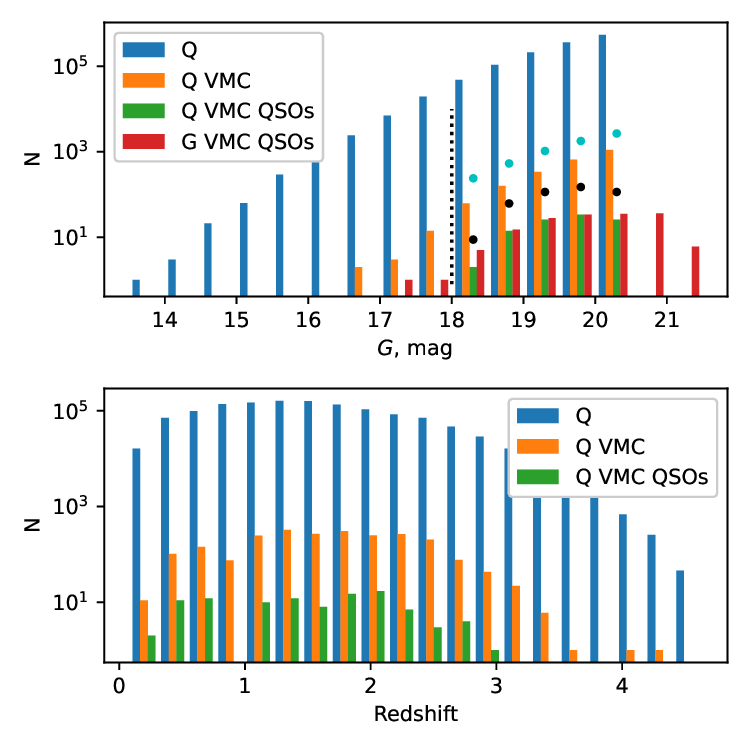} \\
\caption{{\it Top:} Luminosity functions of quasars from Quaia and
the VMC survey in $G$ band:
blue -- for all Quaia quasars (1.3 million objects);
orange -- Quaia quasars with VMC counterparts with quasar-like colors
(2347 objects);
red -- spectroscopically confirmed VMC quasars from this paper and from
\citet{2016A&A...588A..93I} with {\it Gaia} counterparts (161 objects);
green -- same as red, but excluding the 5 objects from the bright SAAO
sample and those that fall below the Quaia $G$=20.5\,mag limit;
black dots -- projected output for a VMC survey full quasar sample
(450 objects);
cyan dots -- confirmed Quaia quasars scaled down from the full sky
to the VMC survey area ($\sim$6300 objects), assuming 10\,\% loss in
the Zone of avoidance, in reality 7386 Quaia quasars fall in the VMC
survey footprint.
{\it bottom:} Redshift distribution:
blue -- for all Quaia quasars (1.3 million objects);
orange -- Quaia quasars with VMC counterparts with quasar-like colors
(2347 objects);
green -- Quaia quasars with spectroscopically confirmed quasars from our
VMC selection (161 objects).
See Sect.\,\ref{sec:results} for details.}\label{fig:quaia_VMC_LFs}
\end{figure}

\begin{figure}[h!]
\centering
\includegraphics[width=9.0cm]{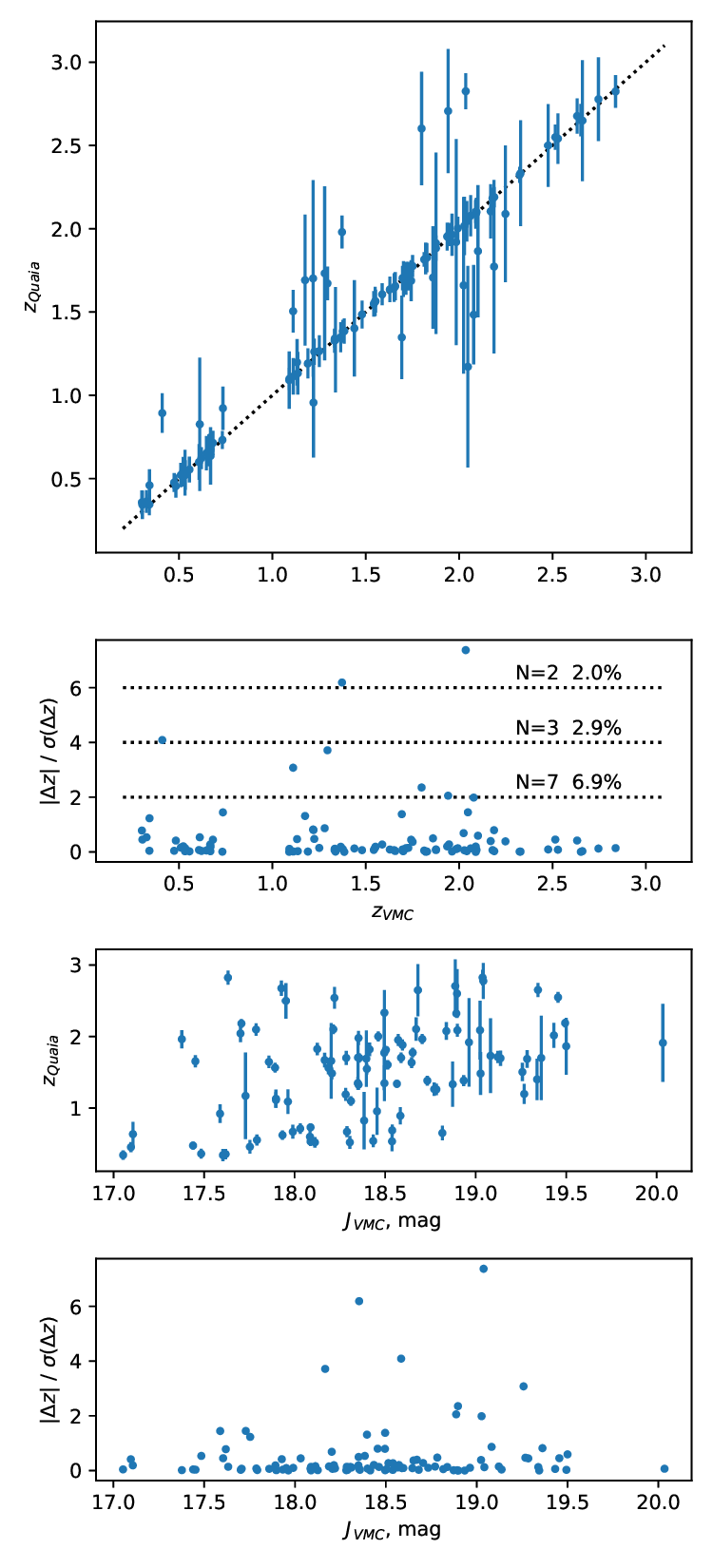} \\
\caption{Comparison of the redshifts from Quaia and from our VLT
spectra (top) and absolute values of the differences as function of
redshift in units of sigma (bottom). The labels above each dashed
line indicate the numbers of quasars above these lines and the
corresponding fractions in percentages. See Sect.\,\ref{sec:results}
for details.}\label{fig:quaia_VMC_z}
\end{figure}

Detections for some of our targets -- all confirmed quasars -- are
reported in the literature:
Bridge 3\_4 109g and Stream 2\_1 174g in UV with {\it GALEX}
\citep[Galaxy Evolution Explorer;][]{2007ApJS..173..682M};
SMC 4\_5 060 and SMC 6\_3 310 in X-ray with {\it XMM-Newton}
\citep[X-ray Multi-Mirror Mission;][]{2013A&A...558A...3S};
SMC 6\_3 310 and SMC 4\_2 071g are listed as new X-ray identified
active galactic nuclei candidates by \cite{2019A&A...622A..29M},
also based on {\it XMM-Newton}, and
SMC 4\_2 071g and again SMC 4\_5 060 were detected in X-ray too,
but with {\it Chandra} \citep{2008MNRAS.383..330M}.
\citet{2012ApJ...747..107K} identified LMC 8\_3 039 as a QSO
candidate from a combination of light curves and multicolor
criteria and our spectrum confirms it.
Summarizing, only a handful of our confirmed QSOs have so far been
detected in X-ray, because of the sparse coverage with sensitive
observations. eROSITA, the soft X-ray instrument on board the
{\it Spectrum-Roentgen-Gamma} ({\it SRG}) mission
\citep{2021A&A...647A...1P} completely covered the region of the
Magellanic Clouds during its all-sky surveys. A preliminary
investigation of the eROSITA catalogue derived from the first
all-sky survey data \citep{2024A&A...682A..34M} reveals a
detection rate up to 30\%, depending on matching criteria. More
detailed work on the X-ray properties of QSOs behind the
Magellanic System is in progress.

\subsection{Completeness}

Controlling the completeness is not a main goal of this paper -- we
aim to increase the number of confirmed quasars spending a minimal
amount of observing time -- but to estimate it nevertheless, we turn
to the sample of nearly 1.3 million quasars (with $G$$<$20.5\,mag;
Fig.\,\ref{fig:quaia_VMC_LFs}, top panel, blue), with secure
redshifts from the low-resolution {\it Gaia} spectra from
\citet[Quaia;][]{2023arXiv230617749S}. This catalog suffers from its
own incompleteness and contamination issues, but the spectroscopic
nature of the confirmation, the large number of quasars and the full
coverage of the VMC survey footprint makes it the most suitable for
this purpose.

First, we determine the number of quasars that our search should
find, that will also be accessible to Quaia.
To compare the output of our spectroscopic confirmation campaign from
this work and from \citet{2016A&A...588A..93I} with the Quaia sample
we scale up our 102 VLT confirmed quasars with Quaia counterparts,
with the inverse fraction of the observed tiles. Between this work
and \citet{2016A&A...588A..93I} we have followed up 25 out of 110 VMC
survey tiles (the seven bright additional objects followed up at the
SAAO are scattered across the entire VMC survey footprint and are
ignored here). Extrapolating over the entire VMC survey we scale up
these 102 VLT confirmed quasars with $G$ in the overlapping range
18--20.5\,mag (to be discussed further) by an area ratio factor of
4.4, arriving at $\sim$450 VMC survey quasars that we may be able to
confirm, applying the same selection and following the same observing
strategy (black dots on Fig.\,\ref{fig:quaia_VMC_LFs}, top). This
number must be compared with the expected number of Quaia quasars
within the VMC survey footprint, over an identical magnitude interval.

On one hand, our estimates are optimistic, because the tiles in this
study are located in the outermost regions in the system where the
crowding and confusion are not as problematic as in the innermost
regions; on the other, it is known that not all quasars are variable
and would meet the variability criterion of \cite{2013A&A...549A..29C}.
Furthermore, the observations were taken under highly variable weather
conditions, because our program was designed as a poor weather filler
and we only followed up the brightest candidates, to optimize the
telescope time.

To obtain $G$ band magnitudes for our candidates and confirmed
quasars we cross-identified them with the {\it Gaia} DR3 main catalog
\citep{2022yCat.1355....0G}, within a radius of 0.35\arcsec. The
results for the former are reported in Table\,\ref{tab_campl_sample},
and for the latter -- in the red histogram Fig.\,\ref{fig:quaia_VMC_LFs},
top panel, for 161 objects (quasars confirmed at the 1.9-m SAAO are
excluded from statistical considerations). It appears 42 ($\sim$24\,\%)
of them can not ever have Quaia counterparts because they fall below
its $G$=20.5\,mag limit, and the bright Quaia quasars fall in the VMC
saturation regime. Here and further the matching radius was
set to make any spurious contamination unlikely: we repeated the
cross-identification with a modified coordinates, increasing the Dec
by 10\arcmin\ and found only 5 matches, three of which are with
separations $>$0.9\arcsec\ and the two others at 0.25--0.35\arcsec\.
This keeps the spurious contamination at $\sim$0.1\,\% level.

The saturation of brighter sources in the VMC is the reason why our
quasars are relatively faint. Bright quasars are important for studies
like those of intervening absorption lines but they constitute
only a small fraction: integrating the Quaia luminosity function above
$G$=16.5\,mag we find that quasars that have not a single VMC survey
counterpart constitute $\sim$0.1\,\% of the entire Quaia sample, and
brighter than $G$=18.0\,mag, which is the brightest limit of our
confirmed quasars (black dashed line on the upper panel) -- only
$\sim$2.3\,\%.

Second, we determine the number of Quaia quasars that are accessible
to the VMC survey based quasar search and follow up.
We cross-identified Quaia quasars with the VMC source catalog
(as of DR6): 7386 Quaia quasars fall within the VMC survey
footprint; 6923 have matches within 0.35\arcsec\ radii, but only for
3149 we have photometry in all three bands of sufficiently high
quality -- with no contamination or error flags and errors $<$0.15\,mag;
2347 matches remain if our color selection criteria are applied. Here
we consider all objects, regardless of their apparent $G$ band
magnitude. However, for a proper comparison we should exclude objects
that are too bright and would saturate in the VMC -- they would reside
in the bright bin with $G$$<$$<$17.5--18\,mag, as seen in
Fig.\,\ref{fig:quaia_VMC_LFs}. Integrating the Quaia luminosity function
above this limit indicates that only $\sim$2\,\% of Quaia quasars are
lost, so we ignore them for the purpose of this comparison.

Finally, our expected yield of 450 confirmed quasars -- within the
Quaia apparent magnitude range -- from the entire VMC survey constitute
$\sim$6\,\% of all Quaia quasars. This fraction increases to
$\sim$19\,\% of nearly 2350 Quaia quasars that match our color criteria.
We speculate that these low fractions come from the weak variability of
most quasars, that our survey data can not recognize. This is consistent
with our preliminary investigation of the light curves of Quaia quasars
with multi-epoch counterparts in the VMC survey: it suggests that
the vast majority of Quaia quasars are indeed not variable enough to
meet our minimum light curve slope criterion. This warrants further
investigation and will be reported in another paper.

\begin{figure*}[!ht]
\centering
\includegraphics[width=18.4cm]{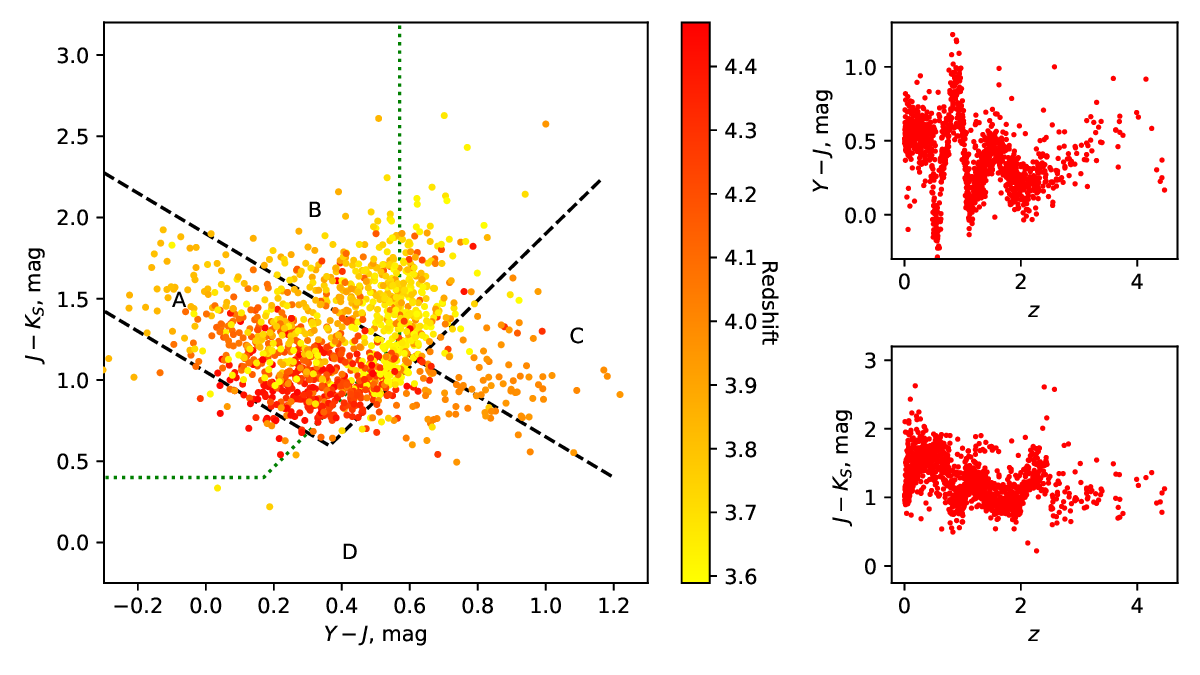} \\
\caption{{\it Left:} Color--color diagram of all spectroscopically
confirmed QSOs with VISTA/VIRCAM photometry. For the notation of
regions see Fig.\,\ref{fig:CCD}.
{\it Right:} $Y$$-$$J$ and $J$$-$$K_S$ colors as a function of
redshift for all spectroscopically confirmed QSOs with VISTA/VIRCAM
photometry. The variations are driven by strong emission lines
entering and exiting the band passes of individual filters.
}\label{fig:CCD_redsift}
\end{figure*}

This result suggests that the complete list of our candidates presented
in Table\,\ref{tab_campl_sample} must contain only a small fraction of
all quasars within the VMC footprint. However, the success rate of our
new spectroscopic follow up is $\sim$90\,\% ($\sim$91\,\% for VLT and
$\sim$71\,\% for 1.9-m), implying that nearly nine for every ten
candidates in the list are likely to be confirmed as quasars, if a
similar spectroscopic follow up is carried out.



The distribution of redshifts for the entire Quaia (blue), the Quaia
quasars with VMC counterparts (orange) and our spectroscopically
confirmed quasars with Quaia counterparts (green) are shown in the
lower panel of Fig.\,\ref{fig:quaia_VMC_LFs}. Our sample is dominated
by lower redshift quasars than Quaia, probably because we tend to
select for follow up the brightest candidates in each tile, minimizing
the observing time per object.
\citet{2023arXiv230617749S} make a considerable effort to verify the
quality of their redshifts. A comparison with the redshift derived
from our 102 VLT spectra suggest major disagreements occur in only a
few percent of the cases (Fig.\,\ref{fig:quaia_VMC_z}), lending extra
credibility to the Quaia catalog. The disagreeing redshifts are not
concentrated towards the faintest objects but occur over a range of
apparent magnitudes, and usually their Quaia redshift errors are
significantly larger than for other objects with similar magnitudes.

\subsection{Color-redshift relations}

We explored if the diagnostic color--color diagram that we use to
select candidates can help to constrain their redshift. To expand
the statistics we added to the VMC QSOs the QSOs from the latest 
catalog of \citet[][13 edition]{2010A&A...518A..10V} that have been 
observed with some other VISTA/VIRCAM surveys: 
VHS \cite[][5719 objects]{2013Msngr.154...35M},
VIDEO \cite[][339 objects]{2013MNRAS.428.1281J}, and 
VVV \cite[][5 objects]{2010NewA...15..433M}.
The color--color diagram coded by redshift is shown in the left 
panel of Fig.\,\ref{fig:CCD_redsift}. The lowest redshift QSOs do
cluster in a locus at $Y$$-$$J$$\sim$0.45--0.65\,mag and 
$J$$-$$K_S$$\sim$1.1--1.8\,mag, but the more distant scatter over 
the entire diagram. The reason for this behavior can be understood
from the panels on the right that show how the two colors vary 
with redshift -- there are sharp color changes as various more 
prominent emission lines enter or exit the bandpasses of individual
filters. Therefore, this diagram has the potential of separating 
only the nearest QSOs.

\section{Summary}\label{sec:summary}

We spectroscopically confirmed 136 QSOs within the footprint of the 
Magellanic system. They were selected from their near-IR colors and 
variability from the ESO VMC public survey. The uniform VMC
observations, spanning nearly 8\,yr proved a reliable resource for
QSO selection, because nearly 90\,\% of the observed candidates were
quasars. However, a comparison with the Quaia catalog indicated that
our selection recovers only 6-19\,\% of quasars identified from the
{\it Gaia} low-resolution spectra. The fraction depends on whether
the magnitude range, the quality of VMC survey photometry and the
candidate colors are considered. It appears the fraction is relatively
low, because most quasars are not sufficiently variable to meet our
variability criterion. Therefore, our quasar candidate list is far
from complete, but the candidates on it are quasars with a high degree
of certainty. The variability is an important quasar identification
tool, especially for radio-quiet of them. Finally, we report a list of
3609 candidates that meet our criteria, encompassing the entire VMC
survey footprint. Based on the previous statistics we expect that
nearly 90\,\% of them would be confirmed as quasars, should they be
subjected to a spectroscopic follow up similar to the one described
here.

\begin{acknowledgements}
This paper is based on observations made with ESO telescopes at the
La Silla Paranal Observatory under program IDs 092.B-0104(A),
098.B-0229(A) and 099.B-0204(A).
We have made extensive use of the SIMBAD Database at CDS (Centre de
Donn\'ees astronomiques) Strasbourg, the NASA/IPAC Extragalactic
Database (NED) which is operated by the Jet Propulsion Laboratory,
CalTech, under contract with NASA, and of the VizieR catalog access
tool, CDS, Strasbourg, France.
We thank Lisa Crause for efficient support at the SAAO telescope.
JEMC acknowledges STFC studentship.
\end{acknowledgements}

\bibliographystyle{aa}
\bibliography{vmc_qso_114}


\appendix

\section{Finder charts}\label{app:finder_charts}

\begin{figure*} 
\centering
\includegraphics[width=18.5cm]{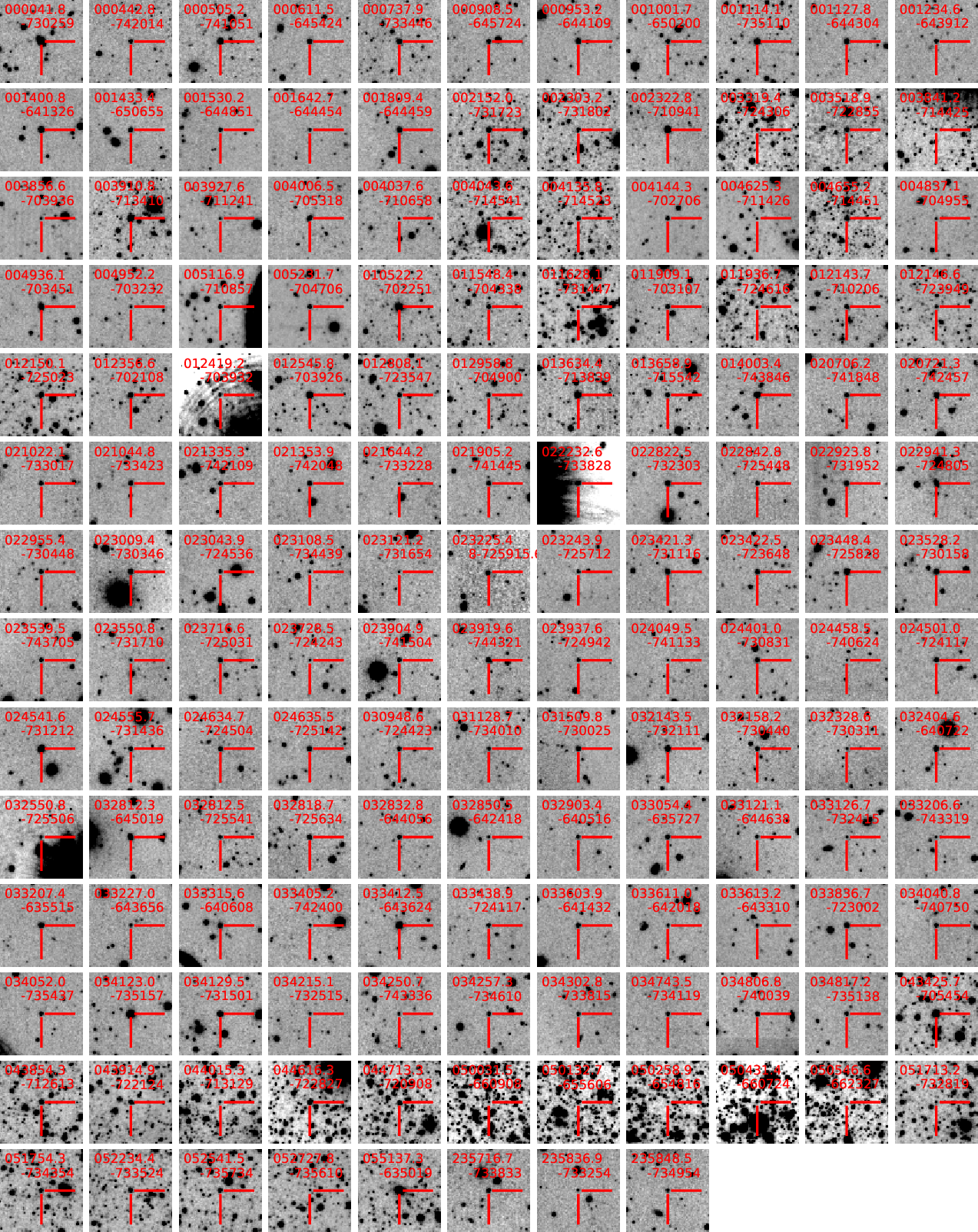} \\
\caption{Finding charts ($Y$ band, 1$\arcmin$$\times$1$\arcmin$)
for all $151$ objects (crosses) with follow up spectroscopy,
sorted by right ascension. North is at the top and east is to the
left.}\label{fig:finders}
\end{figure*}

VMC Survey finder charts of the objects considered here are shown
in Fig.\,\ref{fig:finders}.

\newpage

\section{Complete VMC quasar candidate sample}\label{app:compl_sample}

Table\,\ref{tab_campl_sample} lists 3609 objects that match our
color and variability selection criteria. The comparison with
Quaia catalog indicates that we identify about 7\,\% of their
$G$$\leq$20,5\,mag quasars (see Sect.\,\ref{sec:results}).

A SIMBAD search within 5\arcsec\ from the VMC positions yielded
117 matches:
\begin{itemize}
\item 97 are known quasars, active galaxies, X-ray/radio/blue-UV
sources that may be consistent with galactic activity, or
candidates to any of these classes;
\item 16 are classified as stars, the long period variables and
young stellar objects constituting the largest consistent groups
with 5 and 3 entries, respectively;
\item 2 are galaxies;
\item 2 are objects of unknown nature.
\end{itemize}

\begin{table*} 
\caption{A complete list of quasar candidates in the VMC survey.
Identifiers, coordinates, VMC survey magnitudes with their errors,
number of epochs and covered time span in years are listed. The
$G$ magnitudes and errors of {\it Gaia} DR3 counterparts selected
with a matching radius of 0.35\arcsec\ are also listed. They are
available for 1249 candidates). Only part of the table is shown
for guidance, the entire dataset is given in the electronic edition
only.}\label{tab_campl_sample}
\begin{center}
\begin{small}
\begin{tabular}{@{}l@{ }c@{ }c@{ }c@{ }c@{ }c@{ }c@{ }c@{ }c@{ }c@{ }c@{ }r@{ }c@{ }r@{ }r@{}}
\hline\hline
VMC ID~~~~~~~~~~~~~~~~~~~~~~~~~~~~~~~~~~~~ & \multicolumn{2}{c}{~~~~~$\alpha$~~~~~(J2000)~~~~~~$\delta$~~~~~} & $G$ & $\sigma_G$ & $Y$ & $\sigma_Y$ & $J$ & $\sigma_J$ & $K_S$ & $\sigma_{K_S}$ & Slope~~ & $\sigma_{\rm Slope}$ & N, & ~~$\Delta$T, \\
                                         & (h:m:s) & (d:m:s)                                                 &~(mag)~&~(mag)&~(mag)~&~(mag)&~(mag)~&~(mag)&~(mag)~&~(mag)& & & ep. & yrs\\
\hline
VMC 00:38:25.82$-$75:02:58.6 & 00:38:25.82 & $-$75:02:58.6 &        &       & 19.800 & 0.045 & 19.296 & 0.042 & 17.724 & 0.027 & $-$0.00011 & 0.00003 &  48 & 6.16 \\
VMC 00:41:56.95$-$75:39:04.6 & 00:41:56.95 & $-$75:39:04.6 & 20.004 & 0.008 & 19.336 & 0.032 & 19.216 & 0.039 & 17.866 & 0.029 &    0.00018 & 0.00003 &  48 & 6.16 \\
VMC 00:33:01.86$-$75:44:18.1 & 00:33:01.86 & $-$75:44:18.1 &        &       & 19.850 & 0.047 & 19.587 & 0.054 & 18.434 & 0.044 &    0.00017 & 0.00005 &  47 & 6.16 \\
VMC 00:42:22.73$-$75:32:55.8 & 00:42:22.73 & $-$75:32:55.8 &        &       & 19.917 & 0.049 & 19.519 & 0.050 & 18.378 & 0.042 & $-$0.00015 & 0.00005 &  46 & 3.11 \\
VMC 00:42:33.89$-$74:48:23.9 & 00:42:33.89 & $-$74:48:23.9 & 20.369 & 0.010 & 19.458 & 0.035 & 19.337 & 0.043 & 18.100 & 0.034 &    0.00014 & 0.00001 &  87 & 8.08 \\
VMC 00:33:38.37$-$75:44:43.8 & 00:33:38.37 & $-$75:44:43.8 &        &       & 19.715 & 0.043 & 19.192 & 0.040 & 17.578 & 0.025 & $-$0.00010 & 0.00003 &  50 & 6.16 \\
VMC 00:35:07.73$-$75:20:34.2 & 00:35:07.73 & $-$75:20:34.2 &        &       & 19.773 & 0.044 & 19.444 & 0.047 & 18.609 & 0.049 &    0.00015 & 0.00004 &  43 & 6.16 \\
VMC 00:35:06.88$-$75:23:14.1 & 00:35:06.88 & $-$75:23:14.1 & 18.960 & 0.006 & 18.216 & 0.015 & 17.941 & 0.017 & 17.079 & 0.019 & $-$0.00014 & 0.00002 &  51 & 6.16 \\
VMC 00:44:23.42$-$75:35:55.1 & 00:44:23.42 & $-$75:35:55.1 &        &       & 19.929 & 0.049 & 19.518 & 0.050 & 18.260 & 0.038 &    0.00020 & 0.00004 &  46 & 3.11 \\
VMC 00:35:38.66$-$75:06:45.3 & 00:35:38.66 & $-$75:06:45.3 & 19.777 & 0.011 & 18.369 & 0.017 & 18.246 & 0.020 & 16.743 & 0.015 & $-$0.00012 & 0.00002 &  51 & 6.16 \\
\hline
\end{tabular}
\end{small}
\end{center}
\end{table*}

\end{document}